%% file: template.tex
\documentclass[preprint,journal]{vgtc}            

\onlineid{1702}

\vgtccategory{Research}

\vgtcpapertype{please specify}

\title{
\textcolor{black}{
Sensitivity to Redirected Walking Considering Gaze, Posture, and Luminance
}
}

\author{%
  \authororcid{Niall L.\ Williams}{0000-0002-0273-883X},
  \authororcid{Logan C.\ Stevens}{0000-0002-3893-3012},
  \authororcid{Aniket Bera}{0000-0002-0182-6985}, and
  \authororcid{Dinesh Manocha}{0000-0001-7047-9801}
}

\authorfooter{
  \item
  	Niall L. Williams, Logan C. Stevens, and Dinesh Manocha are with the University of Maryland, College Park.
  	E-mail: \{niallw$\vert$lsteven7$\vert$dmanocha\}@umd.edu
  \item
  	Aniket Bera is with Purdue University.
  	E-mail: ab@cs.purdue.edu.
}

\abstract{%
  We study the correlations between redirected walking (RDW) rotation gains and patterns in users' posture and gaze data during locomotion in virtual reality (VR).
    To do this, we conducted a psychophysical experiment to measure users' sensitivity to RDW rotation gains and collect gaze and posture data during the experiment.
    Using multilevel modeling, we studied how different factors of the VR system and user affected their physiological signals.
    In particular, we studied the effects of redirection gain, trial duration, trial number (i.e., time spent in VR), and participant gender on postural sway, gaze velocity (a proxy for gaze stability), and saccade and blink rate.
    Our results showed that, in general, physiological signals were significantly positively correlated with the strength of redirection gain, the duration of trials, and the trial number.
    Gaze velocity was negatively correlated with trial duration.
    Additionally, we measured users' sensitivity to rotation gains in well-lit (photopic) and dimly-lit (mesopic) virtual lighting conditions.
    Results showed that there were no significant differences in RDW detection thresholds between the photopic and mesopic luminance conditions.
}


\teaser{
  \centering
  \includegraphics[width=0.75\linewidth, alt={todo.}]{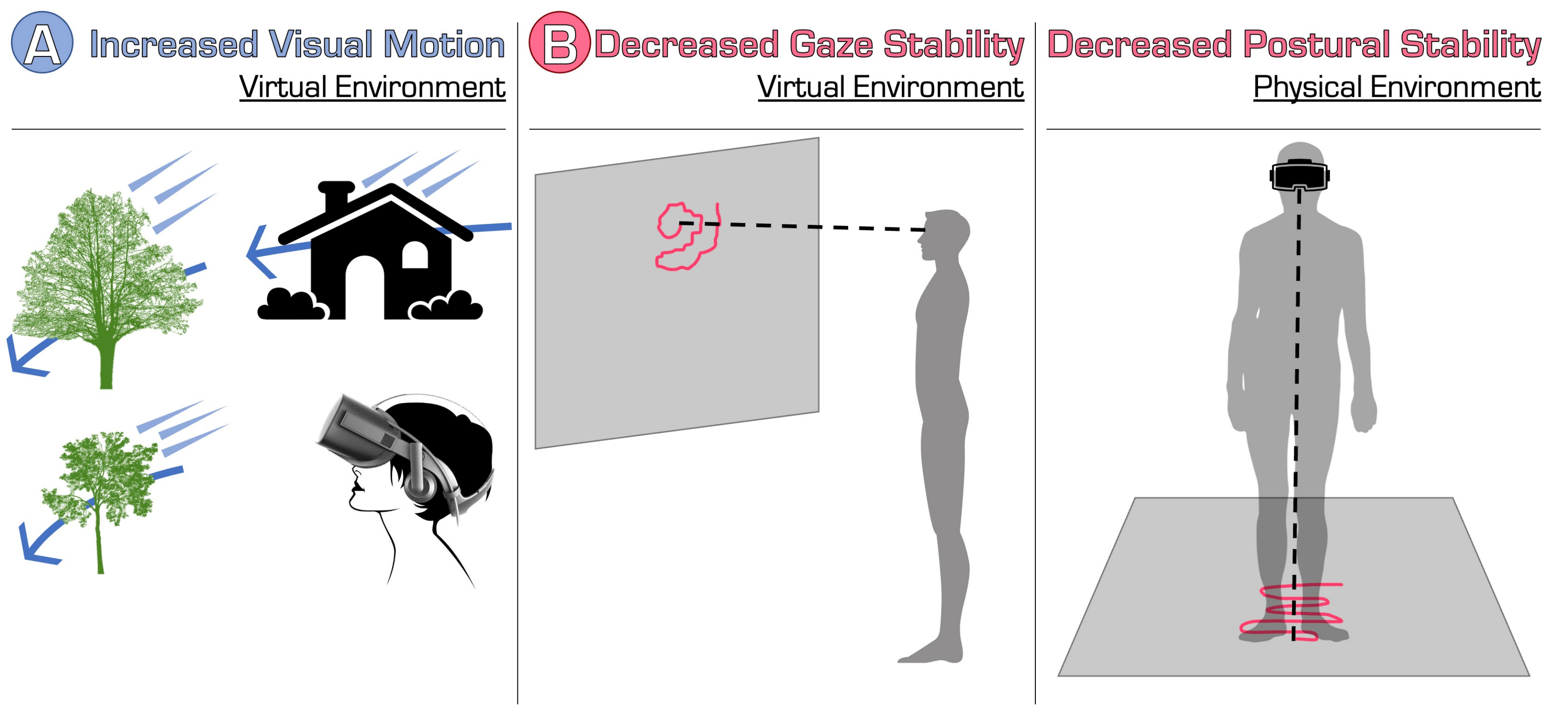}
  \caption{A visualization of our experiment paradigm and the properties of physiological signals that we found to be correlated with scene motion during redirected walking (RDW). \textbf{(A)} We conducted a psychophysical experiment in which participants completed a rotation task across hundreds of trials, with different amounts of additional scene motion injected into the virtual environment during the rotation. Participants reported on whether or not they perceived the additional injected motions and we computed their visual sensitivity to these motions. \textbf{(B)} Our analyses revealed that as the speed of injected motions increased, the stability of participants' gaze (left) and posture (right) \textit{decreased}. These results show, for the first time, a direct correlation between the strength of redirection (injected visual motion gains) and physiological signals.}
  \label{fig:teaser}
}

\graphicspath{{figs/}{figures/}{pictures/}{images/}{./}} 

\usepackage{tabu}                      
\usepackage{booktabs}                  
\usepackage{lipsum}                    
\usepackage{mwe}                       

\usepackage{mathptmx}                  

\usepackage{amsmath}
\usepackage{enumitem}
\usepackage{soul}
\setstcolor{red}

\begin{document}


\maketitle

\input{introduction}

\input{background}

\input{methods}

\input{results}

\input{discussion}
\input{conclusion}

\acknowledgments{%
    The authors thank Evan Suma Rosenberg for helpful discussions about the experiment design. We also thank Jeanine Stefanucci and Matt Gottsacker for advice and help regarding the data analysis. Niall L. Williams was supported in part by the Link Foundation Modeling, Simulation, \& Training Fellowship.
}

\bibliographystyle{abbrv-doi-hyperref}

\bibliography{template}

\end{document}

%% file: introduction.tex
\section{Introduction}
\label{sec:intro}





Natural walking in virtual environments (VEs) is important for providing users with a comfortable and immersive virtual reality (VR) experience \cite{steinicke2013human,di2021locomotion}.
Redirected walking (RDW) is a promising natural locomotion interface that allows users to explore VEs that are larger than the available physical environment (PE) \cite{razzaque2001redirected,razzaque2005redirected}.
RDW works by adding imperceptible rotations and translations (controlled by \textit{gains}) to the virtual camera's trajectory through the VE as the user moves around in the PE.
In order for RDW to be \textcolor{black}{most} effective, it is \textcolor{black}{usually preferred} that these added movements are small enough to remain imperceptible to the user.
If they are not, it is common for the user to feel symptoms of simulator sickness or have difficulty controlling their locomotion in the VE \cite{nilsson201815,steinicke2009estimation}.
To avoid this, researchers have measured the maximum strength of redirection that can be injected before users can reliably perceive it; the weakest RDW gain that is perceptible to the user is known as the \textit{perceptual threshold} (or \textit{detection} threshold), and it is generally advised that gains should not exceed this threshold value in order to avoid causing discomfort for the user \cite{steinicke2009estimation}, \textcolor{black}{although more recent work has shown that users do tolerate some supra-threshold redirection \cite{rietzler2018rethinking}}.

A significant amount of research has been done to study how RDW thresholds change according to different system configurations and using different measurement methods.
One major takeaway that can be drawn from prior work is that perceptual thresholds for RDW are highly variable depending on system and user factors: thresholds change depending on gender \cite{williams2019estimation,nguyen2020effect}, field of view \cite{williams2019estimation}, level of embodiment \cite{nguyen2020effect}, amount of exposure to RDW \cite{bolling2019shrinking}, audio feedback \cite{gao2020visual}, and from user to user \cite{hutton2018individualized}.
\textcolor{black}{As scientific progress advances, the technical capabilities of HMDs get better; for example, recent HMDs are supporting features such as high dynamic range (HDR) imaging \cite{matsuda2022realistic,matsuda2022hdr}, varifocal optics \cite{zhao2023retinal,saeedpour2024perceptual}, and retinal resolution displays \cite{zhao2023retinal}.
Thus, in order to develop more robust and effective RDW systems, it is important that we continue to evaluate how these different system factors may affect users' sensitivity to redirection.
In this paper, we evaluate and measure how display luminance affects users' sensitivity to rotation gains.}

Typically, RDW thresholds are measured using a psychophysical experiment, which often takes multiple hours for one participant, can be very repetitive and tiring for users, and different psychophysical fitting methods can yield different results under the same experimental conditions or may be sensitive to lapses in perception \cite{wichmann2001psychometric1,wichmann2001psychometric2}.
The high variability of RDW thresholds is not compatible with the psychophysical methods used to measure them---small changes in the system configuration may yield different thresholds that would warrant re-running the long psychophysical calibration process in order to ensure that the applied redirection gains remain below the user's detection thresholds.
Thus, in order for RDW to become a more usable locomotion interface in real-world consumer settings, it is crucial that we develop better methods for \textit{quickly} and \textit{accurately} estimating the imperceptibility of RDW gains during virtual locomotion.
Physiological signals may be one potential solution to this problem.
Many researchers have shown a link between users' perceptual experiences in VR (e.g., sense of presence \cite{meehan2002physiological} and jitter perception \cite{lutwak2023user}) and their physiological signals (e.g., gaze signals \cite{orlosky2019using}, galvanic skin response \cite{islam2019measuring}, EEG signals \cite{krokos2022quantifying}, and postural stability \cite{murata2004effects}).
Similarly, researchers in the human vision community have shown that gaze behavior can be informative for estimating an observer's level of engagement with the scene content \cite{ranti2020blink}.
However, despite the growing evidence that physiological signals can be useful for inferring about a user's subjective experience and internal state, researchers have yet to demonstrate a direct correlation between RDW gains and any physiological signals.
\textcolor{black}{Therefore, in this work we also measure how physiological signals (gaze and posture data) are correlated with redirection strength and how system factors (in our case, display luminance) affect these signals.
We are interested in studying display luminance in particular because of the advent of HDR HMDs and prior research that shows that luminance can have an effect on motion perception, gaze patterns, and postural control.}

\textbf{Main Results:} In this work, we take a first step towards a more efficient and accurate estimation of RDW gain perceptibility by investigating the correlations between RDW gain strength and patterns in users' physiological signals \textcolor{black}{across a variety of system factors}.
We conducted a psychophysical experiment to measure users' perceptual thresholds for RDW rotation gains \textcolor{black}{at both well-lit (photopic) and dimly-lit (mesopic) display luminance levels,} and recorded physiological signals (gaze and posture data) during the experiment.
The first main contribution of our work is that we measured how rotation gain thresholds change as a function of the luminance of the virtual content.
We measured perceptual thresholds in photopic and mesopic 
\textcolor{black}{display luminance} conditions to observe how the perceptibility of rotation gains changes depending on the light level of the virtual content being viewed.
The second main contribution of our work is an initial investigation into the correlations between RDW rotation gain strength and patterns in gaze and posture data.
To improve the applicability of our results, we chose to study gaze and posture data because these are physiological signals that are readily available in most modern consumer HMDs; other signals typically require the integration of additional sensors.
We limited our study to rotation gains since they allow for a wider range of gains to be applied before users begin to perceive them, compared to curvature and translation gains \cite{steinicke2009estimation}.
Our results showed that gaze and postural stability are significantly correlated with the properties of the RDW system, and that detection thresholds do not change significantly in photopic compared to mesopic luminance conditions.
\textcolor{black}{Our results suggest that future research on self-motion perception in VR should strongly consider the role of multisensory integration, and lay the initial groundwork for developing non-invasive physiological metrics for feelings of sickness during redirection.}
\textcolor{black}{Specifically}, we observe that:
\begin{itemize}[itemsep=0.1pt]
    \item Detection thresholds for RDW rotation gains were not significantly different between the well-lit (photopic) and poorly-lit (mesopic) virtual lighting conditions.
    
    \item There were small but significant correlations between rotation gain strength and physiological signals of posture and gaze. Specifically:
    \begin{itemize}[itemsep=0.3pt]
        \item Each unit (1) increase in rotation gain was correlated with, on average, a postural sway increase of $5.096$ cm, gaze velocity increase of 
        \textcolor{black}{$0.9034$}$^\circ / s$, and a saccade frequency increase of 
        \textcolor{black}{$5.796$} per trial.
        
        \item Each additional trial completed (i.e., time spent in VR under the effects of RDW) was correlated with, on average, a gaze velocity increase of 
        \textcolor{black}{$0.0280$}$^\circ /s$.
        
        \item Each additional second spent under the effects of RDW per trial (i.e., trial duration) was correlated with a gaze velocity decrease of 
        \textcolor{black}{0.358}$^\circ /s$, a blink frequency increase of 
        \textcolor{black}{$0.388$} per trial, and a saccade frequency increase of 
        \textcolor{black}{$2.430$} per trial.
        \item An interaction between gain and trial duration had a significant effect on number of blinks
        (\textcolor{black}{$-0.324$}) and on gaze velocity
        (\textcolor{black}{$0.9030$}).
    \end{itemize}

\end{itemize}

%% file: background.tex
\section{Background and Related Work}
\label{sec:background}

\subsection{Redirected Walking Thresholds}
Redirected walking (RDW) is a locomotion interface that enables users to explore VEs that are larger than their PE \cite{razzaque2001redirected}.
It works by adding imperceptible rotations and translations to the virtual camera's trajectory as the user moves around in the physical space, creating a mismatch between the user's physical and virtual movements.
The magnitude of these injected rotations and translations is determined by parameters called \textit{gains}, and the smallest gain that is reliably perceptible to a user is known as their \textit{perceptual threshold}.
Rotation gains modify the amount that the user turns in place, translation gains modify how far the user travels while walking, and curvature gains cause users to walk on curve physical trajectories while following straight virtual trajectories.
Ideally, the gains applied during locomotion are smaller than the perceptual thresholds in order to mitigate the likelihood that the user feels symptoms of simulator sickness or breaks in presence \cite{steinicke2009estimation}.

Applying perceptually comfortable thresholds requires us to estimate the user's perceptual thresholds, and this is typically done using psychophysical experiments which often take multiple hours to complete per participant \cite{gescheider2013psychophysics}.
The first comprehensive study that measured RDW thresholds was conducted by Steinicke et al. \cite{steinicke2009estimation}.
Since then, many researchers have measured RDW thresholds in a wide range of contexts and have shown that HMD field of view \cite{williams2019estimation}, walking speed \cite{neth2012velocity,nguyen2018individual}, level of embodiment in VR \cite{nguyen2020effect_emobdiment}, the rate of change of gains \cite{congdon2019sensitivity}, gender \cite{williams2019estimation,nguyen2018individual,nguyen2018individual}, the amount of exposure to RDW \cite{bolling2019shrinking}, and individual differences \cite{nguyen2018individual,hutton2018individualized} have an effect on thresholds.

These results show that RDW thresholds are complex and nuanced: thresholds depend on user factors, device factors, VE context, and amount of user experience.
As VR technology improves and increases in popularity, VR devices will gain new features and capabilities, and users will engage with VR in a wider range of (physical and virtual) contexts.
Thus, in order to help make RDW a practical technology that does not require hours of calibration before it can be used in every new scenario, our work investigates the use of physiological signals for estimating users' ability to perceive redirection.
Furthermore, to test the reliability of these signals in a variety of different configurations, we study how thresholds change according to the luminance level of the display.
We chose to study thresholds as a function of luminance because it is a common visual feature of virtual content that is likely to vary widely for different virtual experiences, and since we expect luminance to become a more-relevant feature in future devices with the advent of high-dynamic range VR \cite{matsuda2022realistic}.
Furthermore, prior work in motion perception \cite{guo2021effects,grossman1999perception,nakamura2013effects} provides evidence that luminance has an effect on an observer's perception of motion, which has implications for users' perception of RDW gains in VR.


\subsection{Physiological Signals of Users' Internal State}
\label{subsec:background_physio_signals}


In addition to thresholds being highly dependent on system and user factors, another challenge with understanding user sensitivity to RDW is the fact that psychophysical experiments are often very time consuming and expensive \cite{owen2021adaptive,watson1983quest,wichmann2001psychometric1}.
Psychophysical experiments often take multiple hours and require users to complete a repetitive, tedious task in order for researchers to estimate their sensitivity to the stimulus in question.
To mitigate this problem, researchers have developed different techniques for sampling the stimulus parameter space to more efficiently converge to the user's detection threshold (e.g. PEST \cite{taylor1967pest}, QUEST \cite{watson1983quest}, AEPsych \cite{owen2021adaptive}), but these methods can still be sensitive to lapses in judgement or may require careful calibration to maximize the likelihood of convergence.
Furthermore, traditional psychophysics requires users to first perceive the stimulus (possibly multiple times) and then answer a question about what they perceived, which is an intrusive process that requires its own dedicated calibration session.
To help mitigate these constraints, researchers have shown that physiological signals can be used to draw conclusions about a user's subjective feelings of comfort or other aspects of their internal state such as level of engagement.

\subsubsection{Gaze Stability}
Prior work has shown that gaze movements, and in particular metrics for gaze stability, are correlated with symptoms of motion sickness (MS) in users.
Ujike et al. \cite{ujike2004effects} found that torsional eye movements had a significant correlation with symptoms of MS for rotations about the roll axis.
Similarly, Hemmerich et al. \cite{hemmerich2020visually} found a small but significant correlation between MS and eye movements (specifically, scan-path length and dispersion).
Brument et al. \cite{brument2021studying} showed that rotation gains had an effect on gaze dispersion and offset.
Bouyer et al. \cite{bouyer1996torso2} had participants repeatedly perform a ``torso rotation'' movement in which they shifted their gaze side-to-side between two targets placed at either end of their horizontal peripheral vision, which required coordinated head, eye, and torso rotations (similar to the rotation movement participants completed in our rotation gain discrimination task).
Bouyer et al. found that gaze stability (measured by participants' ability to consistently focus their gaze on a target) decreased as feelings of MS increased.
Inspired by those works, we evaluate the correlations between redirection strength and participants' average gaze velocity, where gaze velocity is used as a proxy for gaze stability since we did not instruct participants to fixate on a particular target in the virtual environment.
Additionally, we measure correlations between redirection strength and saccade and blink frequency since these are gaze metrics that are readily available in eye-tracked HMDs.

\subsubsection{Postural Stability}
Posture control and \textcolor{black}{retinal} optic flow are tightly linked to locomotion control \cite{lappe1999perception,kelly2008visual,matthis2022retinal,hollman2007does,janeh2017walking}.
Prior work has shown that optic flow modulates locomotion such that users alter their locomotion behavior to better match the perceived optic flow \cite{eikema2016optic,prokop1997visual}.
In VR, researchers have used measures of postural stability to estimate users' level of simulator sickness (SS).
Chardonnet et al. \cite{chardonnet2017features} showed that the spread of the center of gravity and the shape of postural sway in the time domain can serve as indicators of SS.
Soffel et al. \cite{soffel2016postural} demonstrated how a VR HMD can be used to measure postural stability without the need for a balance board or force plates.
Riccio et al. \cite{riccio1991ecological} proposed a theory that postural instability is the source of MS that user experience.
Numerous studies have been conducted that provide some empirical evidence to support this theory \cite{stoffregen1998postural,baltzley1989time,stoffregen2010stance,tanahashi2007effects,stoffregen2008motion}, though some studies have found a lack of relationship between postural stability and feelings of MS \cite{cobb1998static,dennison2017cybersickness,warwick1998evaluating}.
Mostajeran et al. \cite{mostajeran2024analyzing} and Bruder et al. \cite{bruder2015cognitive} found a correlation between the strength of curvature gains and lateral posture sway.
Matsumoto et al. \cite{matsumoto2018biomechanical} found that curvature gains can alter to in/out during gait, Cortes et al. \cite{tirado2019analysis} found that translation gains can impact users' step length and frequency and Janeh et al. \cite{janeh2017walking} found variations in gait parameters depending on the mapping between physical and virtual motions.
Brument et al. \cite{brument2021studying} also showed that rotation gains had an effect on how quickly users rotated their head to follow a moving target.
Given the evidence from prior work that shows a link between postural sway and MS and gait properties, we chose to study correlations between postural sway and redirection gain strength since SS is a common side effect of RDW.




\subsection{Luminance, \textcolor{black}{Perception, \& Action}}
\label{subsec:luminance_motion}
\textcolor{black}{In this section we summarize research on the interactions between luminance and motion perception, gaze behavior, and posture control.}

\subsubsection{\textcolor{black}{Motion Perception}}
Researchers in the human vision community have studied the degree to which an observer's ability to detect motion is affected by the luminance of the visual content.
Sara et al. \cite{sara2017effect} found that observers had decreased motion detection performance in low-light conditions, while Takeuchi et al. \cite{takeuchi2000velocity} showed that velocity discrimination thresholds vary depending on luminance level.
For illusory motion in static stimlui, Hisakata et al. \cite{hisakata2008effects} showed that motion perception varied as a function of retinal illuminance.
Interestingly, Billino et al. \cite{billino2008motion} found that detection thresholds for biological motion were the same for photopic and scopotic conditions, despite the majority of prior research showing that motion perception performance generally decreases as luminance decreases \cite{yoshimoto2016motion}.
Overall, the literature on motion perception in different light levels shows that perception of motion changes according to changes in luminance, which motivated us to study how users' sensitivity to RDW gains changes as a function of the luminance of the virtual content shown on the HMD.
Indeed, with the introduction of high dynamic range VR HMDs \cite{matsuda2022realistic}, we believe that it is important for us to gain a better understanding of motion perception in VR according to the light level of virtual content.

\subsubsection{\textcolor{black}{Gaze Behavior}}
\label{subsubsec:lumniance_gaze_background}
\textcolor{black}{
There is a significant amount of research in the vision science community that studies gaze behavior during whole-body rotations in both light and dark illumination conditions (which is very similar to the experiment we conduct in the present study).
During whole-body rotations, the visual system employs the vestibulo-ocular reflex (VOR) to generate compensatory eye movements that allow observers to fixate in space while moving \cite{goebel1991headshake}.
When a head-fixed target is present, observers are able to eliminate these VOR movements and maintain a stable gaze on the visual fixation target \cite{van2010eye}.
Additionally, when the observer can see a full visual field of motion during rotation (e.g., in a well-lit environment) with no visually-static fixation target, they employ the optokinetic nystagmus (OKN) response to maintain a stable retinal image and reduce blur during rotation \cite{abadi2002mechanisms}.
However, when the observer rotates in complete darkness (i.e., little-to-no visual motion and no visual target present), they employ a vestibular nystagmus response \cite{abadi2002mechanisms}.
In either case, eye movement is characterized by rhythmic pattern of a slow phase and a fast phase of movement, wherein the observer's gaze remains stable in space until during the slow phase until the fast phase triggers a brief saccade causes the eye to jump to a new orientation \cite{abadi2002mechanisms}.
Thus, during whole-body rotations in both light and dark conditions, observers tend to exhibit cyclic gaze patterns characterized by a period of fixation followed by a saccade.
This suggests that gaze behavior in the case of whole-body rotations may not be notably different in light and dark conditions.
}

\subsubsection{\textcolor{black}{Posture Control}}
\textcolor{black}{
Rugelj et al. \cite{rugelj2014influence} studied postural stability in both young (early 20s) and elderly (early 70s) adults when completing a stationary standing task.
Their study found that mesopic lighting conditions yielded significantly worse postural stability in both young and elderly adults.
Kinsella-Shaw et al. \cite{kinsella2006effects} conducted a similar study and found similar results: low light produced more postural instability (with a greater effect on the older participants).
Additionally, they found evidence suggesting that the observer's visual contrast sensitivity was a more important predictor of postural stability.
Focusing only on healthy older women, Brooke-Wavell et al. \cite{brooke2002influence} also found that anteroposterior sway increased in mesopic lighting conditions and was worst when participants' eyes were closed.
It is important to note, however, that these studies only looked at stationary standing tasks and not tasks involving active locomotion.
Huang et al. \cite{huang2017reduced} found that gait stability in young (early 20s) participants decreased as the light level decreased during a treadmill walking task.
However, Naaman et al. \cite{naaman2023young} only found such an effect in middle-aged and elderly populations.
Overall, these studies provide support for the notion that as the illumination of an observer's surroundings decreases, the amount of visual feedback available to support the appropriate perception-action feedback loop to maintain stable posture is diminished, which yields more postural sway.}

%% file: methods.tex
\section{Experimental Methodology}
\label{sec:methods}

In order to \textcolor{black}{most-}effectively utilize RDW, the injected rotations and translations should lie below the user's detection thresholds so as to avoid making the user feel sick.
However, measuring these detection thresholds is a tiring and repetitive process that often entails multiple hours of psychophysics.
Furthermore, it is well-known that a user's sensitivity to RDW can change depending on various system and user factors such as field of view \cite{williams2019estimation}, amount of exposure to RDW \cite{bolling2019shrinking}, and individual differences in perception \cite{hutton2018individualized}.
This poses a problem since the dynamic nature of RDW perceptibility implies that thresholds should be estimated frequently (as different factors influence a user's sensitivity change during the virtual experience), but current methods for estimating sensitivity to RDW often require multiple hours or, at best, on the order of $\sim \!\! 20 - 30$ minutes for adaptive methods \cite{wichmann2001psychometric1}.
Our motivation with this work is to investigate to what degree the strength of redirection gains is correlated with different physiological signals of the user, in hopes that such insights could be used to estimate a user's sensitivity to RDW in real time via online analysis of physiological signal patterns, thus bypassing the long psychophysical estimation process that has traditionally been used for RDW.

To achieve this, we ran a psychophysical study in which we estimated participants' RDW gain detection thresholds while also collecting different physiological signals (posture and gaze data).
After collecting the data, we studied the correlations between redirection strength and different metrics derived from participants' physiological data.
We limited our study to rotation gains since they allow for larger magnitudes of redirection to be applied \cite{steinicke2009estimation} (giving us a wider range of gain values with which we can test for correlations).
Additionally, we chose to study participants' posture data and gaze data since these are two biosignals that are readily available in most modern VR devices without requiring external devices, and since there is a large amount of prior work (\autoref{subsec:background_physio_signals}) to suggest that redirection gains will be correlated with patterns in users' physiological data.

Note that in this work, we are \textit{not} directly measuring a correlation between physiological signals and a user's subjective level of comfort (i.e., tolerance for redirection gains); we only measure the existence of any correlations between redirection gain strength and posture/gaze data.
Since we are not aware of any prior work that has established a direct correlation between redirection gain strength and physiological signals, the first step to using physiological signals as an alternative to psychophysics is to confirm that the strength of redirection gains is indeed correlated with patterns in a user's biosignals.
Once such a correlation has been established, follow-up works should study to what extent these signals can serve as a direct indicator of a user's subjective level of comfort during locomotion.

Additionally, in an attempt to better understand the different factors that may affect sensitivity to RDW \textcolor{black}{and patterns in physiological signal data}, we measured users' sensitivity to rotation gains in both bright (photopic) and dark (mesopic) luminance conditions.
Considering the gamut of different virtual experiences users may find themselves in VR, it is not unlikely that users will interact with virtual environments that have different ranges of luminance values (e.g. visual content for horror games tend to have low luminance, while job simulators are more likely to have bright visual content akin to daytime luminance).
Prior work in human perception has shown that observers can have differing sensitivity to visual motion\textcolor{black}{, gaze behavior, and postural control} depending on the luminance of the visual stimuli (\autoref{subsec:luminance_motion}).
Considering these facts and the introduction of high dynamic range VR \cite{matsuda2022realistic}, we were motivated to study how sensitivity to redirection changes depending on the luminance of the virtual content.

\subsection{Experiment Design \& Stimuli}
Our experiment was limited to rotation gains because rotation gain experiments require less physical space to conduct and they allow for larger redirection than curvature or translation gains \cite{steinicke2009estimation}.
We had two hypotheses that we aimed to test with our experiment:
\begin{itemize}
    \item[\textbf{H1}] Rotation gains will be significantly correlated with patterns in participants' gaze and posture data.
    \item[\textbf{H2}] Rotation gain thresholds are different in mesopic viewing conditions compared to photopic viewing conditions.
\end{itemize}
For \textbf{H1}, we hypothesized that the strength of redirection would predict changes in participants' posture and gaze data since prior work showed that these biosignals are correlated with feelings of MS, and MS is frequently induced by prolonged exposure to RDW \cite{nilsson201815}.
The intuition behind \textbf{H2} is that prior work \textcolor{black}{has shown various effects of environment luminance on an observer's sensitivity to visual motion, gaze behavior, and postural control (\autoref{subsec:luminance_motion}).} 
Therefore, we hypothesized that the light level of the virtual content would have an effect on users' ability to detect the presence of RDW gains \textcolor{black}{and on their gaze and postural data when redirection is applied}.

To measure RDW thresholds, we conducted an experiment that required users to repeatedly complete a 2-alternative forced choice~(2AFC) task \cite{gescheider2013psychophysics}.
Participants were tasked with rotating their whole body in place where they stand, either left or right (indicated by an arrow near the top of the field of view), until they heard a beep tone that was their signal to stop rotating.
After the participant stopped rotating and maintained their current orientation for 1 second, a green check mark appeared and the trial ended.
If the user rotated too far or in the wrong direction at the start of the trial, a yellow arrow appeared that signaled them to turn in the other direction to complete the trial.
The amount rotated in the virtual environment was always $90^\circ$ ($\pm5^\circ$ so as to avoid requiring infeasibly-precise rotation from the participants), but the amount of physical rotation varied depending on the strength of the redirection gain applied.
After the participant completed the rotation task, the view faded to black and the user was prompted to answer the following question \cite{williams2019estimation,steinicke2009estimation}: \textit{``Was the virtual movement smaller or greater than the physical movement?''} (Response options: Smaller, Greater).
\textcolor{black}{For each experimental block,} the gain applied on each trial was randomly chosen from a set of nine gains $\{ 0.6, 0.7, 0.8, 0.9, 1.0, 1.1, 1.2, 1.3, 1.4 \}$, and each gain was tested 12 times for a total of 108 trials \textcolor{black}{per block (216 trials across the whole experiment)}.
In order to elicit natural eye movements, the direction arrow disappeared shortly after the user began rotating and participants were not given any instruction on where they should look during the rotation trials.


An observer's ability to discern objects in their surroundings changes depending on the amount of light illuminating their environment.
As the light level changes, the eye adapts and vision improves over time in a process called light adaptation.
In particular, when shifting to a low-light environment, it can take up to two hours for the observer to fully adjust to the low-light conditions \cite{bruce2003visual}.
This required us to separate \textcolor{black}{the experiment into two blocks, one with a light and one with a dark display luminance}, so that the participants' eyes could adequately adjust to the light level of either condition.
If light and dark trials were interleaved within the same block, participants would not have enough time to meaningfully adjust their vision to properly perceive the virtual content.
Blocks were separated by at least 12 hours and the order in which participants experienced them was counterbalanced.
In the light condition, the average luminance of the virtual content in the HMD was about $25~\frac{cd}{m^2}$ (photopic vision).
In the dark condition, the average luminance was about $0.4~\frac{cd}{m^2}$ (mesopic vision).
The dark condition was implemented using a shader that uniformly lowered each pixel's brightness value to $2\%$ of its original value.
Pilot tests revealed that it was not feasible to change the luminance of the virtual content to a scotopic level while having the virtual content still visible to the user.

\begin{figure}[!t]
    \centering
    \includegraphics[width=.4\textwidth]{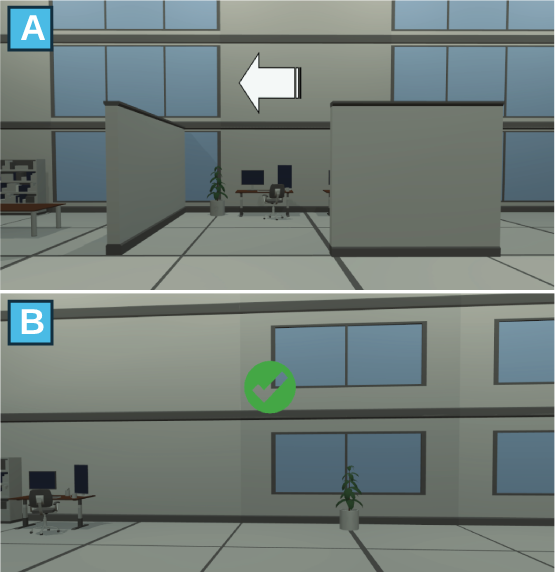}
    \caption{Screenshots of the virtual office environment used in our experiment (during the photopic condition). Ambient office sounds were played to help mitigate the viability of using sounds from the physical environment as a cue for the participant's orientation in the physical space. \textbf{(A)}~The view of the environment that participants saw at the beginning of each trial. The white arrow indicated to the user which direction they should rotate, and this arrow disappeared after they rotated $5^\circ$ from the starting position in the direction of the arrow. \textbf{(B)}~An example view of the environment at the end of a trial. When the user rotated $90^\circ$ in the virtual environment ($\pm 5^\circ$), a beep tone was played that indicated that the user should stop rotating and maintain their current orientation in the environment. After maintaining this orientation for 1 second, a green check mark appeared to indicate that they successfully completed the trial.
    }
    \label{fig:stimuli}
\end{figure}

The virtual environment was an office with some pieces of office furniture placed throughout the environment.
We used this environment for our experiment since it is a familiar setting that would have enough variation in color contrast and texture patterns to elicit sufficient optical flow for adequate motion perception \cite{warren2001optic}. 
An example of the stimuli users saw in the photopic condition is shown in \autoref{fig:stimuli}.
After completing a block, simulator sickness symptoms were recorded using the Kennedy-Lane Simulator Sickness Questionnaire (SSQ) \cite{kennedy1993simulator}.
\textcolor{black}{Participants received a $\$10$ Amazon gift card after completing the first block, and a $\$40$ Amazon gift card after completing the second block (\textit{i.e.,} $\$50$ in total if they completed both blocks of the experiment).}
\textcolor{black}{The experiment and procedure were approved by the authors' Institutional Review Board.}

\subsection{Equipment \& Participants}
Our experiment was implemented using the Unity 2021.3.5f1 game engine and a Meta Quest Pro VR HMD.
The HMD was tethered to a desktop computer (CPU: Xeon 2.2GHz, RAM: 64GB, GPU: dual GTX2080, operating system: Windows 10).
We conducted the experiment in a private lab room with covered windows so that the light level in the room could be carefully controlled.
In the mesopic condition, the lights in the room were dimmed to mitigate any adaptation to scoptic light levels in case the participant took a break and removed the HMD during the experiment.
The room could not be completely darkened because the Quest Pro uses inside-out tracking that requires enough ambient light to be able to track the relative motion of the physical surroundings.
We recruited eleven participants (four female, age $\mu=22.636, \sigma=3.107$) who successfully completed both blocks of the experiment.
\textcolor{black}{Some of our recruited participants did not complete the experiment, either because they did not show up for the second block or requested to end the experiment early due to feelings of sickness.}
Note that for psychophysical experiments, it is common to have a single-digit number of participants since each participant provides a large amount of data and each individual participant is treated as an individual replication of the experiment \cite{smith2018small,huang2023simple}.
That is, the objective of psychophysics is to study the mechanics of the perceptual system, so average population effects are largely uninformative since hypotheses are tested within the individual participants \cite{huang2023simple}.
Participants had normal or corrected-to-normal vision and were at least 18 years old.
The HMD was adjusted to match the participant's interpupillary distance and the eye tracker was calibrated for the user.

%% file: results.tex
\section{Results}
\label{sec:results}

\subsection{Rotation Gain Thresholds}
\begin{figure}[t]
    \centering
    \includegraphics[width=.45\textwidth]{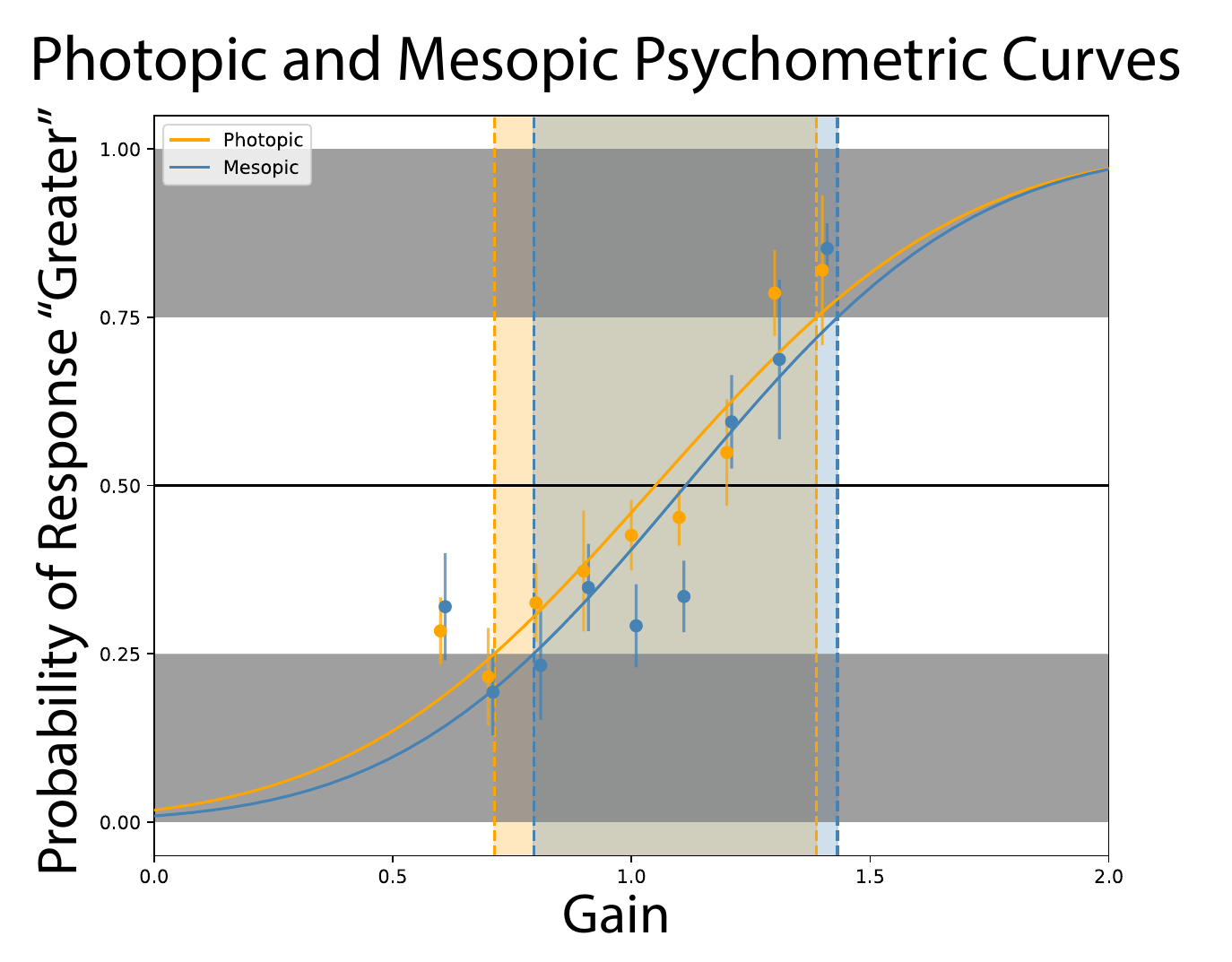}
    \caption{Psychometric curves fit to participants' pooled response data for the photopic (yellow) and mesopic (blue) conditions. The graph shows the average probability of responding \textit{``greater''} to the post-trial question \textit{``Was the virtual movement smaller or greater than the physical movement?''}. The yellow- and blue-shaded regions indicate the estimated range of rotation gains that are usually imperceptible to users (i.e., the 25\% and 75\% detection thresholds). Error bars for each data point denote the standard error. The pooled detection thresholds for photopic and mesopic conditions were similar to values found in prior work that used photopic stimuli, and there were no significant differences between the two conditions. The detection threshold gains shown here are not exactly the same as the average values shown in \autoref{tab:thresholds} since we computed the curves in this plot by fitting a psychometric curve to the \textit{pooled} participant responses, while \autoref{tab:thresholds} computes the average of the curves fit to \textit{individual} participants' responses for each conditions.
    }
    \label{fig:psychometric_curves}
\end{figure}

Rotation gain thresholds were computed by fitting a cumulative Gaussian distribution to participants' average likelihood of responding \textit{``greater''} for a trial.
Threshold estimates are shown in \autoref{tab:thresholds}.
To evaluate the effects of display luminance on perception thresholds, we conducted a 9 (Gain: $0.6:1.4:0.1$) $\times$ 2 (Luminance: photopic, mesopic) ANOVA.
We found that gain had a significant effect on the likelihood of the participant to respond \textit{``greater''} ($F(8, 42)=10.405$, $p<.0001$, $\eta ^2=0.39$), but found no effect of luminance on this likelihood of \textit{``greater''} response ($F(1, 42)=1.405$, $p=0.238$, $\eta ^2=0.00543$).
Additionally, no significant interaction effects between threshold and luminance were found ($F(8, 42)=0.497$, $p=0.857$, $\eta ^2=0.02$).
These results do not support our 
\textcolor{black}{second} hypothesis that detection threshold values would be different in low-light and well-lit viewing conditions.
A plot of the psychometric curves for mesopic and photopic conditions averaged across all participants is shown in \autoref{fig:psychometric_curves}.


\begin{table}[!t]
    \centering
    \begin{tabular}{lllll} 
         ID (luminance) & 25\% & $\mu$ (PSE) & $75\%$ & $\sigma$ \\ \midrule \midrule
        1 (photopic) & 0.689 & 1.06 & 1.43 & 0.552 \\
        2 (photopic) & 0.026 & 0.719 & 1.41 & 1.03 \\
        4 (photopic) & 0.671 & 0.958 & 1.25 & 0.426 \\
        5 (photopic) & 0.911 & 1.13 & 1.35 & 0.328 \\
        6 (photopic) & 0.945 & 1.08 & 1.22 & 0.207 \\
        7 (photopic) & 0.561 & 0.964 & 1.37 & 0.598 \\
        9 (photopic) & 0.862 & 1.04 & 1.22 & 0.266 \\
        13 (photopic) & 0.213 & 1.20 & 2.19 & 1.46 \\
        14 (photopic) & 0.191 & 0.808 & 1.42 & 0.915 \\
        15 (photopic) & 0.747 & 0.943 & 1.14 & 0.292 \\
        16 (photopic) & 1.01 & 1.09 & 1.18 & 0.122 \\ \midrule
        \textbf{Average} & \textbf{0.621} & \textbf{1.000} & \textbf{1.380} & \textbf{0.563} \\ \\ 
        
        ID (luminance) & 25\% & $\mu$ (PSE) & $75\%$ & $\sigma$ \\ \midrule \midrule 
        
        1 (mesopic) & 0.971 & 1.12 & 1.27 & 0.225 \\
        2 (mesopic) & 1.15 & 1.25 & 1.35 & 0.147 \\
        4 (mesopic) & 0.781 & 1.10 & 1.42 & 0.470 \\
        5 (mesopic) & 0.769 & 0.999 & 1.23 & 0.341 \\
        6 (mesopic) & 0.584 & 1.04 & 1.49 & 0.669 \\
        7 (mesopic) & 0.521 & 1.04 & 1.56 & 0.770 \\
        9 (mesopic) & 1.03 & 1.12 & 1.20 & 0.123 \\
        13 (mesopic) & 0.399 & 1.02 & 1.64 & 0.921 \\
        14 (mesopic) & -1.04 & 0.865 & 2.77 & 2.83 \\
        15 (mesopic) & 0.816 & 0.940 & 1.06 & 0.184 \\
        16 (mesopic) & 0.928 & 1.05 & 1.17 & 0.180 \\ \midrule
        \textbf{Average} & \textbf{0.628} & \textbf{1.0495} & \textbf{1.469} & \textbf{0.624} \\
        
    \end{tabular}
    \caption{Individual psychometric fits for each participant in the photopic and mesopic light conditions. We show the point of subjective equality (PSE), the standard deviation of the Gaussian ($\sigma$), and the 25\% and 75\% detection threshold gains. We also show the average for each of these values at the bottom of each sub-table. In general, participants exhibited RDW detection thresholds that were within the ranges found in prior work \cite{langbehn2018redirected}, though there is significant inter-participant variability \cite{hutton2018individualized}. Our results showed no significant differences in detection thresholds between the two lighting conditions.}
    \label{tab:thresholds}
\end{table}

\subsection{Simulator Sickness}
\begin{figure}[t]
    \centering
    \includegraphics[width=.45\textwidth]{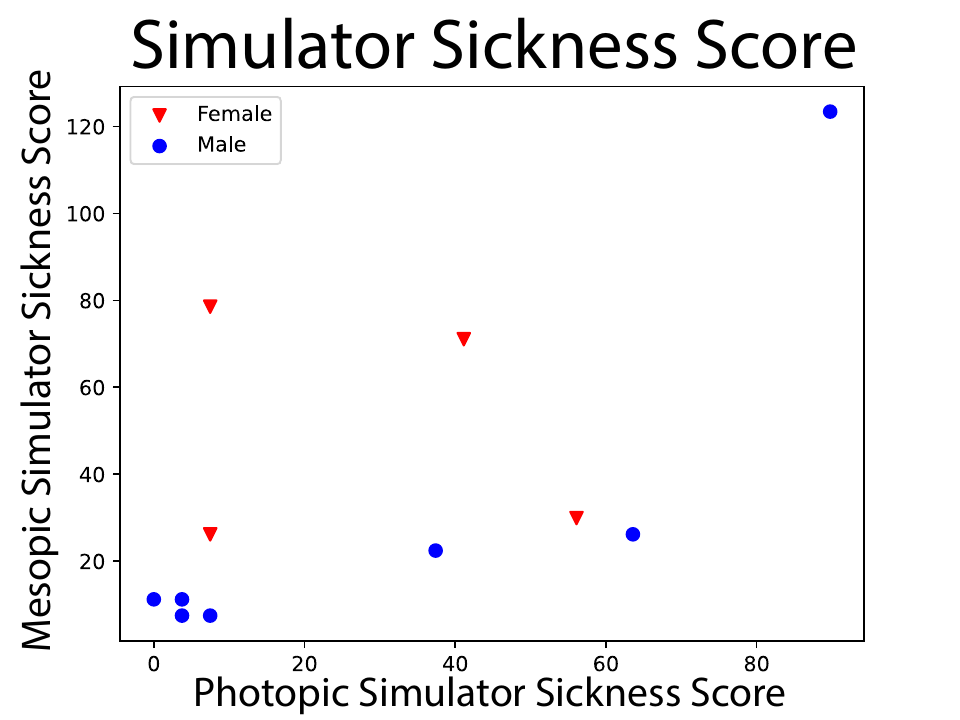}
    \caption{A scatter plot of users' SS scores after the light (photopic) and dark (mesopic) blocks of our experiment. In general, participants exhibited SS levels that are typical of RDW detection threshold experiments. The data belonging to the outlier male participant with the highest SS scores did not show any anomalous patterns, so their data are included in our analyses.}
    \label{fig:ss}
\end{figure}

Simulator sickness values for each participant are shown in \autoref{fig:ss}.
In general, participants' SS values were not abnormal for a rotation gain threshold experiment (Photopic condition: $\mu = 29.421$, $\sigma = 32.478$. Mesopic condition: $\mu = 40.143$, $\sigma = 36.874$).
Participant 5 showed the highest SS levels (89.76 and 123.42 for the photopic and mesopic conditions, respectively) but did not display any concerning symptoms during the experiment.
Inspection of this participant's physiological data and threshold data (\autoref{tab:thresholds}) did not reveal any noteworthy outlying data, so we included their data in our analyses.
A Wilcoxon signed-rank test revealed no significant effect of luminance on SS values ($T=9.0$, $p=0.398$).

\subsection{Postural Data}
\label{subsec:postural_stabililty}

\begin{figure}[!t]
    \centering
    \includegraphics[width=.48\textwidth]{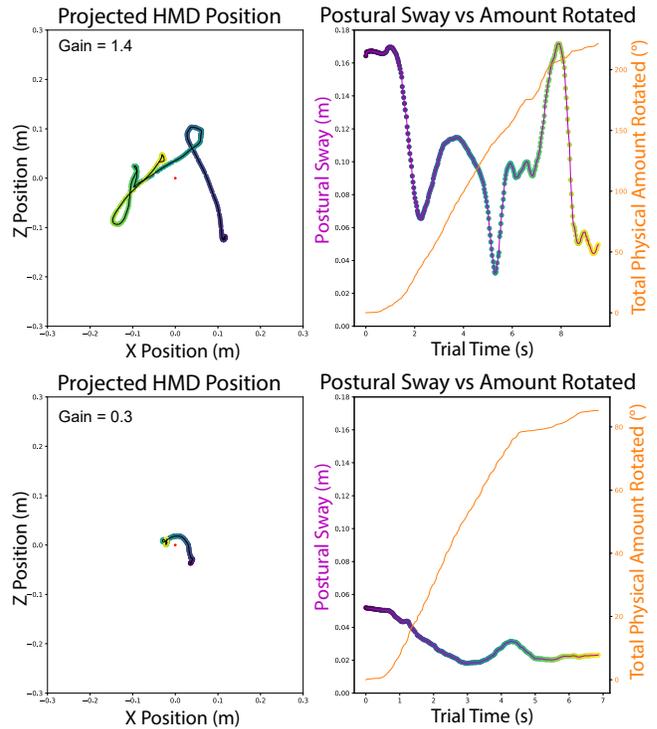}
    \caption{Examples of one participant's posture data for two different trials (each row corresponds to one trial). The left column shows the participant's head position projected onto the ground plane (black curve), with the centroid of their positions at the origin (red dot). For each trajectory point sampled, we compute a proxy for postural sway as the distance between the centroid and the sampled head position (i.e., the distance from each point to the origin). The right column shows the participant's postural sway (purple curve) and total amount rotated in the physical environment (orange curve) across the duration of the trial. The points along the trajectory curves and postural sway curves are colored according to the time in the trial (purple indicates the beginning of the trial, yellow indicates the end of the trial). These plots show that as the gain increases, participants' postural sway also increases---a correlation which was statistically significant (\autoref{subsec:postural_stabililty}).}
    \label{fig:posture_plots}
\end{figure}

First, we compute the average postural stability for each trial.
For each trial, we compute the centroid of the position of the HMD projected onto the floor (i.e., the HMD's 2D position ignoring height) across all trials.
We use this centroid point as a proxy for the user's base position that they would stay in if there were no postural sway.
Then, we compute the distance between this centroid point and the HMD's 2D position for each frame of the trial; this gives a measure of how far the user strays from the central position on any given frame.
Averaging these distances yields our average postural sway metric, where a larger number indicates higher \textit{instability} (more time spent at a distance far away from the centroid point).
Some example plots of one participant's posture data over the course of a trial are shown in \autoref{fig:posture_plots}.
In this paper, all postural sway values are reported in centimeter units.

To analyze the viability of using postural stability as a signal for RDW tolerance, we fitted a multilevel model to the pooled participant trial data using the \texttt{lme4} package and residualized maximum likelihood in R.
By doing this we are able to measure the correlation between independent variables like rotation gain and physiological signals like postural stability.
\textcolor{black}{In the following analyses (\autoref{subsec:postural_stabililty} and \autoref{subsec:gaze_data}) we do not report the Akaike information criterion (AIC) of the models because AIC is a relative metric used when comparing multiple different models on the same data; for each physiological signal of interest we only evaluated one model because the focus of our work is not on developing the most optimal model and because the parameter selection in our models was grounded in our experiment design and hypotheses \cite{sutherland2023practical,vrieze2012model}.}
The model we fit was:
\begin{align}
\begin{split}
avgPosturalSway =& (1 | ID) + gain + gender + trialNum \\
    & + trialDuration \textcolor{black}{+ luminance}\\ & + gain \times trialDuration.
\end{split}
\end{align}
Using this model, we treat the average postural sway as the dependent variable and model how different independent effects may be correlated with postural sway.
Our independent variables are rotation gain, participant ID, participant gender, trial number, and trial duration.
Participant ID is treated as a random variable, while all others are fixed variables.

Results of fixed effects estimates indicated that gain (slope~$=~5.096$, SE~$=~0.467$, $t~=~10.905$, $p~<~0.0001$) was significantly correlated with the average postural sway.
Intuitively, this can be interpreted as meaning that as the gain value increases by 1, the average postural sway increases by $5.096$~cm.
No significant main effects of 
\textcolor{black}{other parameters} were observed.
Random effects analysis revealed very little variability between participants (variance~$=~0.0211$, SD~$=~1.453$), which indicates that the results were not noticeably biased by any outlier participants who completed our experiment (postural sway varied by only $0.0211$~cm from participant to participant, on average).

\subsection{Gaze Data}
\label{subsec:gaze_data}


In general, participants' gaze behavior during trials was characterized by the optokinetic reflex \cite{schweigart1997gaze,lencer2019smooth} and vestibular nystagmus \cite{ivy1929physiology,lencer2019smooth}.
Since vestibular nystagmus is known to be present in both light and dark lighting conditions \cite{graybiel1946role}, we did not separate the analyses based on lighting conditions.
To study correlations between gaze behavior and the strength of redirected walking, we measured how a different gaze-related metrics changed over the course of our threshold experiment.
In particular, we looked at average gaze velocity ($^\circ/s$), number of blinks, and number of saccades.
We classified saccades as any eye movements that exceeded $30^\circ /s$ \cite{van2011defining,dai2021detection}.
Note that since we did not instruct users to fixate on a specific target, we could not directly measure gaze stability (in a manner similar to postural stability in \autoref{subsec:postural_stabililty}).
That is, without knowing exactly what target the user intends to fixate on, we cannot measure to what extent their gaze deviates from said target.
Instead, we studied gaze velocity \textit{only during fixations} as a proxy measure for gaze stability.
Manual inspection of patterns in participants' vestibulo-ocular reflex (VOR) gain did not reveal any noteworthy patterns (the VOR gains during fixations were near 1, as expected) so we did not conduct any further VOR gain analyses.

We fitted a separate multilevel model to each of the three gaze metrics we studied (average velocity, number of blinks, and number of saccades) using the \texttt{lme4} package in R with residualized maximum likelihood.
By modeling average gaze velocity per trial as the dependent variable and participant ID and gender, rotation gain, trial number, and trial duration as independent variables (\autoref{eqn:gaze_vel_model}), we found that gain (slope~$~=~0.9034$, SE~$~=~0.324$, $t~=~5.112$, $p~<~0.0001$), trial number (\textcolor{black}{slope~$~=~0.0280$, SE~$~=~0.00885$, $t~=~3.153$}, $p~<~0.001$), and trial duration (\textcolor{black}{slope~$~=~-0.358$, SE $~=~0.1032$, $t~=~-3.472$}, $p~<~0.0001$) were significantly correlated with average gaze velocity.
Random effects analysis revealed that there was very little difference in average gaze velocity between participants (variance~$~=~0.475^\circ/s$, SD~$~=~0.689^\circ/s$).
\begin{align}
\begin{split}
avgGazeVelocity =& (1 | ID) + gain + gender + trialNum \\
    & + trialDuration + \textcolor{black}{luminance} \\ & + gain\times trialDuration
\end{split}
\label{eqn:gaze_vel_model}
\end{align}

We fit the same model for the number of blinks and saccades per trial.
For number of blinks, we found that trial duration (\textcolor{black}{slope~$~=~0.388$, SE~$~=~0.0318$, $t~=~12.190$}, $p~< 0.0001$), \textcolor{black}{luminance(slope~$~=~-0.643$, SE~$~=~0.158$, $t~=~-4.065$, $p~<~0.001$),} and the interaction between gain and trial duration (\textcolor{black}{slope~$~=~-0.324$, SE~$~=~0.0869$, $t~=~-3.728$, $p~=~0.0002$}) were significantly correlated.
The difference in blink frequency between participants was small (variance $~=~0.898$, SD~$~=~0.948$).
For number of saccades, the fitted model revealed a significant correlation with gain (\textcolor{black}{slope~$~=~-5.796$, SE~$~=~2.233$, $t~=~-2.596$, $p~=~0.00955$}), trial duration (\textcolor{black}{slope~$~=~2.430$, SE~$~=~0.1013$, $t~=~23.980$}, $p~< 0.0001$), and the interaction between gain and trial duration (\textcolor{black}{slope~$~=~0.9030$, SE~$~=~0.273$, $t~=~3.307$, $p~=~0.00097$}).



%% file: discussion.tex
\section{Discussion}
\label{sec:discussion}

\subsection{Detection Thresholds and Sickness Scores}
In general, the simulator sickness scores of our participants were similar to prior work (e.g., \cite{williams2019estimation}).
Additionally, we were able to fit a psychometric curve to all of our participants' data without convergence issues (\autoref{tab:thresholds}).
Note that some of the individual fitted thresholds are somewhat wide, though it is already well-known that sensitivity to RDW can vary widely from person to person \cite{hutton2018individualized,nguyen2018individual}.

In our work, we did not find any significant effect of light level on detection thresholds.
We believe this is likely due to the complex, multimodal nature of self-motion perception since participants' ability to detect rotation gains depends on both visual \textit{and} non-visual cues of self-motion.
In our experiment, users had to determine if there were additional visual motion gains added to their virtual movement and \textcolor{black}{then} \textit{compare the differences} between their visual and non-visual signals of self-motion.
Self-motion perception requires the brain to integrate motion signals from different sensory channels into a single, coherent ``story'' about how the body is moving \cite{chen2017cue,shayman2022multisensory}.
Furthermore, this integration process is known to change depending on the \textit{reliability} of the sensory information \cite{fetsch2009dynamic}, such that less reliable sensory inputs contribute less to the determination of body motion.
Thus, it is 
\textcolor{black}{probable} that participants integrated their signals of self-motion differently for the photopic and mesopic conditions in our experiment, and that these different cue-combinations may have yielded \textcolor{black}{overall} similar performances for rotation perception 
\textcolor{black}{between the two luminance conditions}.
\textcolor{black}{Indeed,} informal interviews with participants revealed that some participants employed a different strategy for the photopic and mesopic conditions.
For example, one participant stated that they found themself relying less on the visual information and more on the duration of each rotation trial in addition to their non-visual motion signals to try to detect the redirection.

Furthermore, we note that our experiment task was a rotation discrimination task.
Prior work in vision science has shown that people are able to accurately perceive rotation magnitudes even in complete darkness \cite{siegler2000self}, which may further complicate the nature of RDW detection in photopic and mesopic conditions.
For these reasons, we believe that it is not necessarily surprising that we did not see any significant differences in detection thresholds between the photopic and mesopic conditions.
\textcolor{black}{
Indeed, based on this non-significant result, we strongly suggest that, moving forward, researchers studying RDW thresholds (and self-motion perception in VR more broadly) should design and conduct multisensory studies that take into account not only the users' visual perception of motion, but also their non-visual perception, too.
Indeed, there is ample evidence in the perception community that multisensory integration plays an important role in self-motion perception (see \cite{greenlee2016multisensory,hou2020multisensory} for comprehensive reviews).
}

\subsection{Postural Data}
In our experiment, we found that participants' postural stability was significantly correlated with the strength of redirection gain applied.
The postural instability theory of motion sickness proposes that postural instability is the main cause of motion sickness \cite{riccio1991ecological}.
\textcolor{black}{Stoffregen} et al. showed that postural instability (i.e., increased postural sway) is associated with symptoms of motion sickness \cite{stoffregen1998postural}.
This finding is in line with our result, which showed that as the redirection gain increased (and thus, the user's surroundings become more visually unstable due to additional rotations), participants' postural sway also increased.
\textcolor{black}{Stoffregen} et al. found that postural sway preceding motion sickness ranged, on average, on the order of 10 cm or less \cite{stoffregen1998postural}, and we found that as the gain increases by one unit, postural sway increases on average by 
\textcolor{black}{5.096} cm.
This adds support to the notion that postural sway can be used as a signal for feelings of discomfort for users during redirection.
\textcolor{black}{We did not find any effect of luminance on postural stability. Although prior work has shown a decrease in postural stability as light levels decrease \cite{rugelj2014influence,kinsella2006effects,brooke2002influence,huang2017reduced}, the effects were less pronounced in younger subjects and our participant pool was biased towards young people.
Furthermore, at least one study \cite{naaman2023young} has shown that lower light levels did not decrease postural stability, so it is possible that the effects of luminance on postural stability in the context of VR and RDW are complex and would require a more focused study than ours to fully understand.
}

\subsection{Gaze Data}
Our results on gaze data during RDW showed that the participant's average gaze velocity, number of blinks, and number of saccades were all significantly correlated with the RDW gain\textcolor{black}{,}
time spent being redirected\textcolor{black}{, or the display luminance}.
We used average gaze velocity as a proxy for gaze stability since participants were not tasked with fixating on any target in particular, so gaze stability with respect to a visual target could not be measured.

\begin{figure}[t]
    \centering
    \includegraphics[width=.4\textwidth]{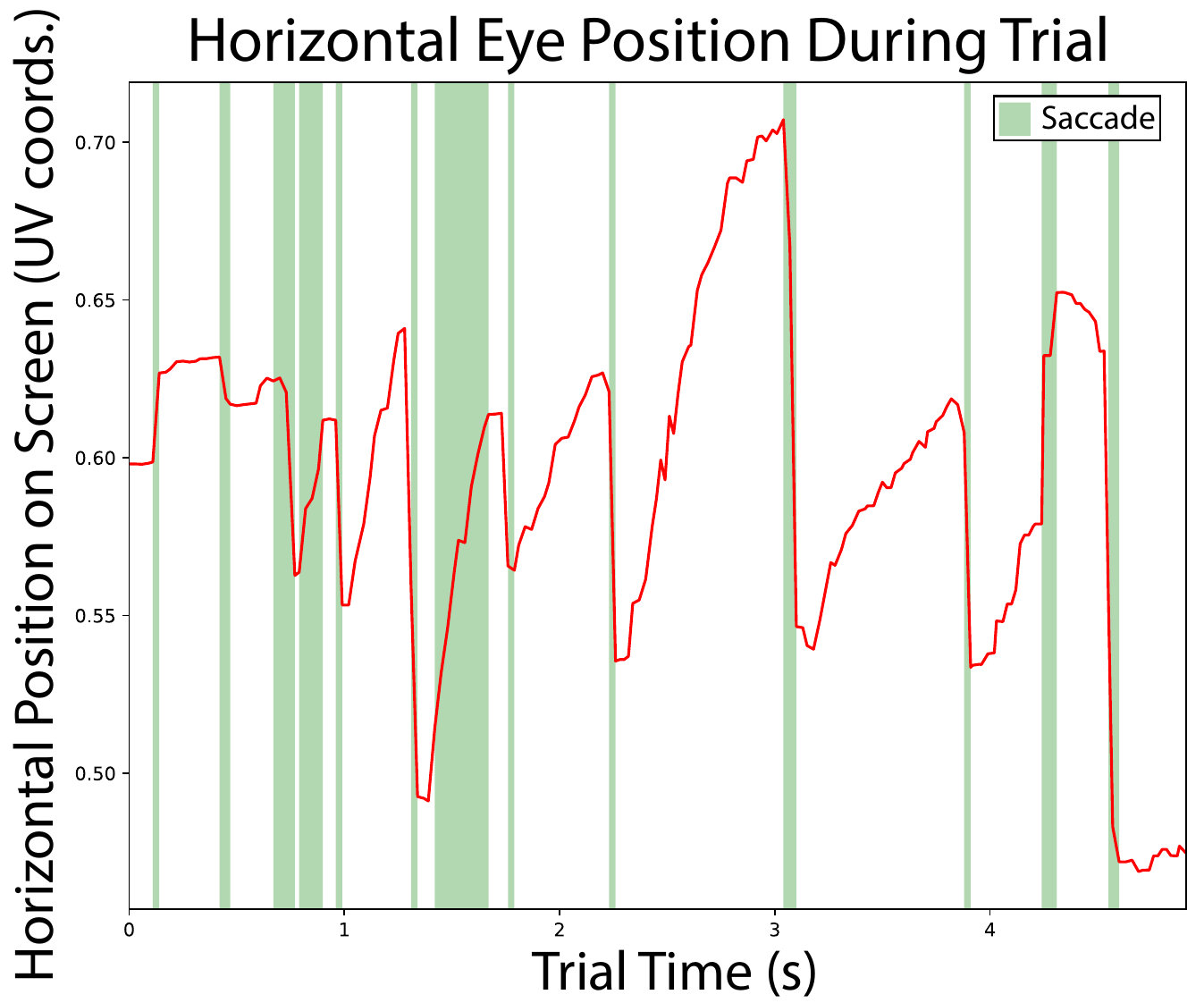}
    \caption{An example of one participant's horizontal eye position (red curve, in UV coordinates of the rendered image) during one trial. Green indicate saccades (gaze velocity above $30^\circ$), which are also identifiable as a very steep slope in the red curve, denoting the eye's horizontal position. The data help to confirm that our participants' gaze behavior was free of abnormalities since this plot shows that gaze behavior was characterized by typical nystagmus responses that are expected in healthy observers during head rotation \cite{abadi2002mechanisms}.}
    \label{fig:horiz_eye_pos}
\end{figure}

As the RDW gain increased (the same physical rotation yielded a larger virtual rotation), we found that participants had higher average gaze velocity.
This make sense when we consider the rotation task and how gaze behavior works during head and body rotations.
In healthy people, nystagmus is induced by body or head rotations as a mechanism for maintaining a stable view of the observer's surroundings.
Optokinetic nystagmus occurs when the observer perceives a large moving field during rotation (as in the photopic condition of our experiment), while vestibular nystagmus occurs during rotations even in darkness \cite{abadi2002mechanisms}.
Gaze position profiles during nystagmus are characterized by a ``saw tooth'' shape as the eye slowly drifts in the direction opposite to the head rotation and then quickly jumps back in the direction of rotation. 
This pattern can be seen in one user's data in \autoref{fig:horiz_eye_pos}.
The frequency of nystagmus is a function of the speed of the head rotation and/or the perceived visual field motion.
Therefore, as the rotation gain increases, the speed of the rotation of the virtual environment increases, which creates more nystagmus-induced saccades that raise the average gaze velocity of the trial.
That is, the larger the redirection gain, the faster the virtual world rotates around the user, which triggers more saccades that increase the average gaze velocity for each trial.

\begin{figure}[t]
    \centering
    \includegraphics[width=.425\textwidth]{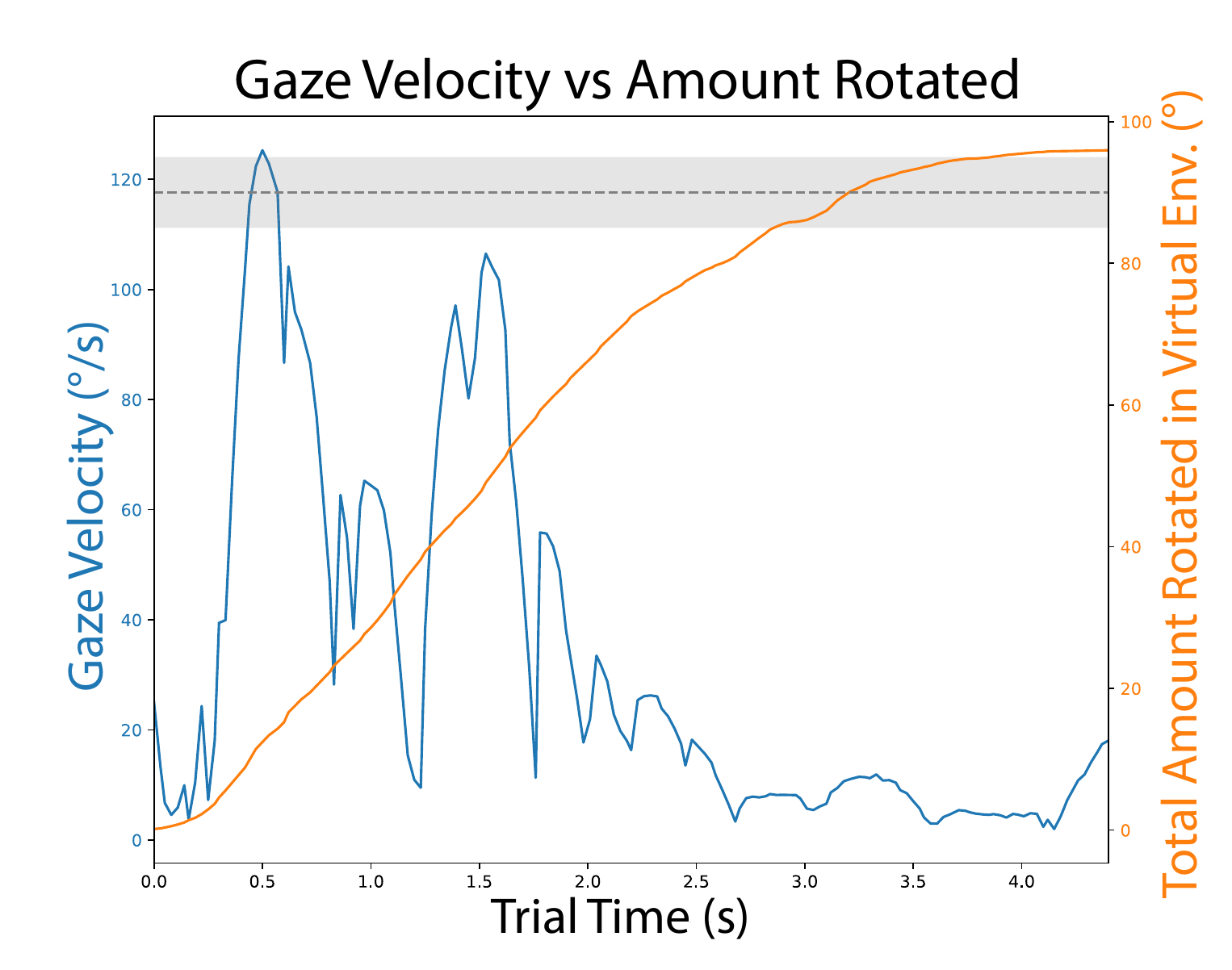}
    \caption{A graph showing a user's gaze velocity (blue) compared to the total amount they have rotated their body during one trial (orange). The gaze velocity curve is characterized by multiple saccades that arise from vestibular nystagmus and the optokinetic reflex. The body rotation curve increases from $0^\circ$ to $\sim \!\! 95^\circ$ over the course of the trial. The grey shaded region is the range $85^\circ - 95^\circ$ (the trial ended if the orange curve lies within this region for 1 s). As the trial progresses, the user's gaze velocity gradually decreases to $0^\circ$ as they wait for the trial to complete after rotating a sufficient amount. 
    }
    \label{fig:gaze_vel_vs_amount_rotated}
\end{figure}

Interestingly, we found a \textit{negative} correlation between trial duration and gaze velocity (\autoref{subsec:gaze_data}).
A closer inspection of the data revealed two patterns that we believe explain this negative correlation.
First, as the participant reaches the end of a trial, their rotation velocity decreases as they try to avoid turning beyond $90^\circ$ in the VE (see the plateau towards the end of the trial time in the orange curves in \autoref{fig:posture_plots} (right column) and \autoref{fig:gaze_vel_vs_amount_rotated}).
As participants slow down their body rotation to complete the trial, their gaze velocity also decreases (see the tail of the blue curve in \autoref{fig:gaze_vel_vs_amount_rotated}) due to the decreased scene motion (fewer nystagmus movements and more smooth vestibulo–ocular reflex movements).
We found that longer trials tend to have longer gaze velocity tails as the participants made small adjustments to their virtual heading in order to complete the trial.
The second factor that we believe contributes to the negative correlation between trial duration and gaze velocity is that of previously-documented slow and fast compensatory phases of eye movements during vestibular stimulation \cite{tanguy2008vestibulo,abadi2002mechanisms}.
When a person's head motion transitions from still to rotating, there is an initial step change in head rotation velocity.
This sudden movement initiates ocular nystagmus that alternates between slow and fast phases: gaze velocity quickly reaches a peak and then slowly decreases exponentially and approaches $0^\circ / \text{s}$ \cite{tanguy2008vestibulo}.
Indeed, this is the pattern seen in the gaze velocity curve (blue) in \autoref{fig:gaze_vel_vs_amount_rotated}---gaze velocity starts off slow ($0~-~0.4$~s), quickly peaks to a maximal value ($0.4~-~0.6$~s) shortly after the user begins turning, and is followed by periodic, smaller spikes in gaze velocity ($0.6~-~2.5$~s) which gradually approach $0^\circ / s$ ($2.5~-~4.5$~s).
Thus, as trial duration increases, it is logical that the participant's average gaze velocity decreases as their gaze velocity profile decreases exponentially.


We found a positive correlation between gain and blink and saccade frequency, which may indicate an increase in simulator sickness symptoms as users spent more time in the experiment \cite{hemmerich2020visually,bouyer1996torso2,dennison2016use,berk1893comprehensive}.
Finally, gaze velocity and blink frequency were correlated with time spent being redirected in VR (trial number) and trial duration.
However, these correlations were \textit{very} small, so they may not be useful signals for users' perception of RDW gains.

\textcolor{black}{
We found that display luminance (photopic versus mesopic) had a significant effect on users' average blink frequency but not any other gaze metrics (average gaze velocity or number of saccades).
We believe this result makes sense given the prior results of studies on gaze behavior and light levels (\autoref{subsubsec:lumniance_gaze_background}).
Nystagmus responses are quite stable in healthy adults, so it is not surprising that participants' saccade rates (and thus their average gaze velocities) were not affected by the luminance condition.
Interestingly, we found that participants blinked more frequently in the mesopic condition.
Recent results in vision science have shown evidence that blinks increase visual sensitivity (i.e., improve visibility) \cite{yang2024eye}.
It may be the case that participants blink more frequently in low-light conditions to take advantage of this improved visibility and counteract the decreased visibility imposed by low luminance.
}

\subsection{Further Considerations\textcolor{black}{, Applications, \& Limitations}}
\subsubsection{\textcolor{black}{Experiment Design Implications \& Limitations}}
Although our results showed some correlations between RDW system factors and users' physiological signals, our study design limits the amount of correlations we were able to study.
We chose to use the method of constant stimuli (MCS) for our experiment because this topic is 
\textcolor{black}{relatively} understudied, so we preferred to have a uniform sampling of the RDW gain parameter space.
This allowed us to study equally how physiological signals varied across the parameter space, as opposed to an adaptive sampling paradigm which will be biased near to the participant's detection threshold.
Since MCS applies gains in a random order, we were unable to test how physiological signals behave near a user's detection threshold since this threshold value is only known after the experiment once we fit a psychometric curve.
As such, we were only able to test for general correlations between RDW gain and physiological signals.
\textcolor{black}{
Measuring detection thresholds requires running a psychophysical experiment, but psychophysics may not be the best paradigm for measuring correlations between physiological signals and visual stimuli.
For example, the temporal dynamics of physiological signals are difficult to measure in a trial-based psychophysical study.
Furthermore, since psychophysical experiments often employ very simplified stimuli and/or experimental tasks, the data collected may not be the most faithful representation of the real-world settings where active monitoring of physiological signals is most useful (see \autoref{subsubsec:applications} for more details).
Therefore, by choosing a traditional psychophysical paradigm for our experiment, it is possible that the reliability of our results on physiological signals is limited.
We suggest that future work that studies physiological correlates of motion perception and comfort in VR use an experimental design that is similar to that of continuous psychophysics instead \cite{bonnen2015continuous}.
}

Furthermore, we chose a randomized order of gains to mitigate learning effects and to minimize the chance that participants feel debilitating levels of simulator sickness.
As a result, it is possible that our data do not capture some patterns in users' physiological data that only manifest after long, continuous exposure to redirection.
With a different study design (such as adaptive staircase or continuous psychophysics), one may be able to better study how physiological signals behave near the detection threshold and during uninterrupted periods of walking.

Prior work has shown differences in detection thresholds depending on gender \cite{williams2019estimation,nguyen2018individual}.
\textcolor{black}{However, in our experiment, since not everyone successfully completed the experiment, our biased participant pool was biased towards male participants (seven male, four female).
Therefore, our ability to draw reliable conclusions about effects of gender on redirection sensitivity or physiological signal patterns is limited.
Future work studying redirection thresholds should ensure parity in participants' gender distribution to mitigate the likelihood of finding biased results \cite{peck2020mind,peck2021divrsify}.
Although psychophysical experiments often employ relatively fewer participants since each new participant is treated as a separate replication of the study \cite{huang2023simple,smith2018small}, we stress that it is still necessary to obtain a large enough sample-size when making cross-population comparisons such as effects of gender or group-level differences in physiological signal patterns.}
\textcolor{black}{Additionally}, we note that while we did see significant correlations, the effects were sometimes small; we believe this is due to the short trial durations and because most of the trials applied a gain that was not near the participant's detection threshold.

In this work, we only measured the correlation between redirection strength and gaze/posture data.
This should not be mistaken as a correlation between \textit{user comfort} and physiological data.
Since we did not continually record any measure of user comfort during the course of the experiment (e.g. after every trial), we cannot claim that there is a direct correlation between physiological signal patterns and user comfort.
Since the strength (and perceptibility) of RDW gains is related to feelings of comfort \cite{rietzler2018rethinking}, it is likely such these correlations exist.

\textcolor{black}{
The HMD used in our experiment was a Meta Quest Pro, which has an eye tracker with a sampling rate of $\sim \! \! 72$ Hz \cite{hou2024unveiling}.
This sampling rate is on the lower end for VR HMDs and for broader research in vision science.
Some eye movements (e.g. microsaccades) are extremely transient and thus require a very fast eye tracker in order to be studied \cite{holmqvist2011eye}.
In our experiment, we only studied fixations and saccades, which can be reasonably well-detected with a $\sim \! \! 72$ Hz eye tracker \cite{holmqvist2011eye}, though it certainly did have a non-negligible effect on the quality of the gaze data we collected.
Despite this limitation, our results showed that we were still able to reproduce the expected gaze behaviors observed in prior experiments with similar experimental tasks (see \autoref{subsec:gaze_data}, \autoref{fig:horiz_eye_pos}, and \autoref{fig:gaze_vel_vs_amount_rotated}).
Thus, we believe that the limitations posed by the HMD's eye tracker is not a major concern for our study.
}

\subsubsection{\textcolor{black}{Applications}}
\label{subsubsec:applications}
\textcolor{black}{
Given that a direct, significant effect of redirection gain strength on physiological signals has been established in this work, we believe this effect opens the door for making RDW more widely-applicable outside of controlled lab settings.
If future research can establish a correlation between a user's physiological signals and their subjective feelings of sickness induced by redirection, researchers and application developers will be able to monitor and analyze a user's physiological data in real time to determine whether or not the applied redirection gains are too strong.
That is, physiological data can serve as a signal that the strength of redirection being applied in the virtual experience should be decreased, to mitigate users' feelings of sickness.
Importantly, developing an understanding of this relationship between physiological data and symptoms of motion sickness will likely allow us to apply redirection gains that are fine-tuned to each user's individual tolerance levels in a wide range of system and user configurations \textit{without} needing to conduct long psychophysical calibration sessions to estimate the user's detection thresholds.
This would allow researchers to develop more robust redirection algorithms that are more effectively capable of providing a comfortable experience for users across a wide variety of virtual experiences.
}

\textcolor{black}{
Furthermore, a better understanding of the relationship between objective physiological signals and redirection gains allows researchers to study the more subtle impacts of added visual motion (introduced either intentionally via RDW or unintentionally via tracking and rendering errors) on users' fine-grained motor control in VR, which has applications for perception-for-action research (e.g. perception for locomotion \cite{muller2023retinal,matthis2022retinal}).
}

%% file: conclusion.tex
\section{Conclusions, Limitations, \& Future Work}
\label{sec:conclusion}

In this work, we investigated correlations between physiological signals (gaze and posture) and RDW gains.
We also studied the impact of photopic and mesopic lighting conditions on users' sensitivity to RDW.
In line with our first hypothesis, we found that postural stability, gaze velocity, blink and saccade frequency were significantly correlated with various factors of the VR environment during redirection (RDW gain, trial duration, and trial number).
Contrary to our second hypothesis, our results showed no significant effect of luminance on RDW thresholds.
Our results contribute to the growing body of knowledge on RDW thresholds and factors that affect it, as well as potential signals for RDW tolerance in the form of physiological signals.
Our work has some limitations.
It is difficult to know exactly which patterns in the physiological data were due to RDW or due to symptoms of simulator sickness that participants will naturally feel as experiment progresses.
However, it is common for RDW to increase users' sickness levels, so in practice it may not be important to disentangle the effects of RDW and simulator sickness on physiological signal patterns.
Additionally, our participant pool was biased toward young people.
Finally, the physiological signals we recorded and analyzed are only a subset of the potentially-available signals that one could record during VR.
We limited our study to posture and gaze data since these are biosignals that are easily obtainable in most modern HMDs, but other signals such as heart rate or skin conductance may be useful correlates as well.
Future work should study correlations between RDW and other signals such as galvanic skin response, pupilometry, and heart rate.

%% file: template.bbl
\begin{thebibliography}{100}

\bibitem{abadi2002mechanisms}
R.~V. Abadi.
\newblock Mechanisms underlying nystagmus.
\newblock {\em Journal of the Royal Society of Medicine}, 95(5):231--234, 2002.

\bibitem{baltzley1989time}
D.~R. Baltzley, R.~Kennedy, K.~Berbaum, M.~Lilienthal, and D.~Gower.
\newblock The time course of postflight simulator sickness symptoms.
\newblock {\em Aviation, Space, and Environmental Medicine}, 60(11):1043--1048, 1989.

\bibitem{berk1893comprehensive}
C.~Berk, C.~Ufuk, and K.~Capin~Tolga.
\newblock A comprehensive study of the affective and physiological responses induced by dynamic virtual reality environments. comp anim virtual worlds 30 (3--4): e1893, 1893.

\bibitem{billino2008motion}
J.~Billino, F.~Bremmer, and K.~R. Gegenfurtner.
\newblock Motion processing at low light levels: Differential effects on the perception of specific motion types.
\newblock {\em Journal of Vision}, 8(3):14--14, 2008.

\bibitem{bolling2019shrinking}
L.~B{\"o}lling, N.~Stein, F.~Steinicke, and M.~Lappe.
\newblock Shrinking circles: Adaptation to increased curvature gain in redirected walking.
\newblock {\em IEEE transactions on visualization and computer graphics}, 25(5):2032--2039, 2019.

\bibitem{bonnen2015continuous}
K.~Bonnen, J.~Burge, J.~Yates, J.~Pillow, and L.~K. Cormack.
\newblock Continuous psychophysics: Target-tracking to measure visual sensitivity.
\newblock {\em Journal of vision}, 15(3):14--14, 2015.

\bibitem{bouyer1996torso2}
L.~Bouyer and D.~Watt.
\newblock “torso rotation” experiments; 2: Gaze stability during voluntary head movements improves with adaptation to motion sickness.
\newblock {\em Journal of Vestibular Research}, 6(5):377--385, 1996.

\bibitem{brooke2002influence}
K.~Brooke-Wavell, L.~Perrett, P.~Howarth, and R.~Haslam.
\newblock Influence of the visual environment on the postural stability in healthy older women.
\newblock {\em Gerontology}, 48(5):293--297, 2002.

\bibitem{bruce2003visual}
V.~Bruce, P.~R. Green, and M.~A. Georgeson.
\newblock {\em Visual perception: Physiology, psychology, \& ecology}.
\newblock Psychology Press, 2003.

\bibitem{bruder2015cognitive}
G.~Bruder, P.~Lubos, and F.~Steinicke.
\newblock Cognitive resource demands of redirected walking.
\newblock {\em IEEE transactions on visualization and computer graphics}, 21(4):539--544, 2015.

\bibitem{brument2021studying}
H.~Brument, M.~Marchal, A.-H. Olivier, and F.~Argelaguet~Sanz.
\newblock Studying the influence of translational and rotational motion on the perception of rotation gains in virtual environments.
\newblock In {\em Proceedings of the 2021 ACM Symposium on Spatial User Interaction}, pp. 1--12, 2021.

\bibitem{chardonnet2017features}
J.-R. Chardonnet, M.~A. Mirzaei, and F.~M{\'e}rienne.
\newblock Features of the postural sway signal as indicators to estimate and predict visually induced motion sickness in virtual reality.
\newblock {\em International Journal of Human--Computer Interaction}, 33(10):771--785, 2017.

\bibitem{chen2017cue}
X.~Chen, T.~P. McNamara, J.~W. Kelly, and T.~Wolbers.
\newblock Cue combination in human spatial navigation.
\newblock {\em Cognitive Psychology}, 95:105--144, 2017.

\bibitem{cobb1998static}
S.~V.~G. Cobb and S.~C. Nichols.
\newblock Static posture tests for the assessment of postural instability after virtual environment use.
\newblock {\em Brain Research Bulletin}, 47(5):459--464, 1998.

\bibitem{congdon2019sensitivity}
B.~J. Congdon and A.~Steed.
\newblock Sensitivity to rate of change in gains applied by redirected walking.
\newblock In {\em Proceedings of the 25th ACM Symposium on Virtual Reality Software and Technology}, pp. 1--9, 2019.

\bibitem{dai2021detection}
W.~Dai, I.~Selesnick, J.-R. Rizzo, J.~Rucker, and T.~Hudson.
\newblock Detection of normal and slow saccades using implicit piecewise polynomial approximation.
\newblock {\em Journal of vision}, 21(6):8--8, 2021.

\bibitem{dennison2017cybersickness}
M.~S. Dennison and M.~D’Zmura.
\newblock Cybersickness without the wobble: Experimental results speak against postural instability theory.
\newblock {\em Applied ergonomics}, 58:215--223, 2017.

\bibitem{dennison2016use}
M.~S. Dennison, A.~Z. Wisti, and M.~D’Zmura.
\newblock Use of physiological signals to predict cybersickness.
\newblock {\em Displays}, 44:42--52, 2016.

\bibitem{di2021locomotion}
M.~Di~Luca, H.~Seifi, S.~Egan, and M.~Gonzalez-Franco.
\newblock Locomotion vault: the extra mile in analyzing vr locomotion techniques.
\newblock In {\em Proceedings of the 2021 CHI Conference on Human Factors in Computing Systems}, pp. 1--10, 2021.

\bibitem{eikema2016optic}
D.~J.~A. Eikema, J.~H. Chien, N.~Stergiou, S.~A. Myers, M.~M. Scott-Pandorf, J.~J. Bloomberg, and M.~Mukherjee.
\newblock Optic flow improves adaptability of spatiotemporal characteristics during split-belt locomotor adaptation with tactile stimulation.
\newblock {\em Experimental brain research}, 234:511--522, 2016.

\bibitem{fetsch2009dynamic}
C.~R. Fetsch, A.~H. Turner, G.~C. DeAngelis, and D.~E. Angelaki.
\newblock Dynamic reweighting of visual and vestibular cues during self-motion perception.
\newblock {\em Journal of Neuroscience}, 29(49):15601--15612, 2009.

\bibitem{gao2020visual}
P.~Gao, K.~Matsumoto, T.~Narumi, and M.~Hirose.
\newblock Visual-auditory redirection: Multimodal integration of incongruent visual and auditory cues for redirected walking.
\newblock In {\em 2020 IEEE international symposium on mixed and augmented reality (ISMAR)}, pp. 639--648. IEEE, 2020.

\bibitem{gescheider2013psychophysics}
G.~A. Gescheider.
\newblock {\em Psychophysics: the fundamentals}.
\newblock Psychology Press, 2013.

\bibitem{goebel1991headshake}
J.~A. Goebel, M.~Fortin, and G.~D. Paige.
\newblock Headshake versus whole-body rotation testing of the vestibulo-ocular reflex.
\newblock {\em The Laryngoscope}, 101(7):695--698, 1991.

\bibitem{graybiel1946role}
A.~Graybiel, B.~Clark, K.~MacCorquodale, and D.~I. Hupp.
\newblock Role of vestibular nystagmus in the visual perception of a moving target in the dark.
\newblock {\em The American Journal of Psychology}, pp. 259--266, 1946.

\bibitem{greenlee2016multisensory}
M.~W. Greenlee, S.~M. Frank, M.~Kaliuzhna, O.~Blanke, F.~Bremmer, J.~Churan, L.~F. Cuturi, P.~R. MacNeilage, and A.~T. Smith.
\newblock Multisensory integration in self motion perception.
\newblock {\em Multisensory Research}, 29(6-7):525--556, 2016.

\bibitem{grossman1999perception}
E.~Grossman and R.~Blake.
\newblock Perception of coherent motion, biological motion and form-from-motion under dim-light conditions.
\newblock {\em Vision research}, 39(22):3721--3727, 1999.

\bibitem{guo2021effects}
X.~Guo, S.~Nakamura, Y.~Fujii, T.~Seno, and S.~Palmisano.
\newblock Effects of luminance contrast, averaged luminance and spatial frequency on vection.
\newblock {\em Experimental Brain Research}, 239:3507--3525, 2021.

\bibitem{hemmerich2020visually}
W.~Hemmerich, B.~Keshavarz, and H.~Hecht.
\newblock Visually induced motion sickness on the horizon.
\newblock {\em Frontiers in Virtual Reality}, 1:582095, 2020.

\bibitem{hisakata2008effects}
R.~Hisakata and I.~Murakami.
\newblock The effects of eccentricity and retinal illuminance on the illusory motion seen in a stationary luminance gradient.
\newblock {\em Vision research}, 48(19):1940--1948, 2008.

\bibitem{hollman2007does}
J.~H. Hollman, R.~H. Brey, T.~J. Bang, and K.~R. Kaufman.
\newblock Does walking in a virtual environment induce unstable gait?: An examination of vertical ground reaction forces.
\newblock {\em Gait \& posture}, 26(2):289--294, 2007.

\bibitem{holmqvist2011eye}
K.~Holmqvist.
\newblock {\em Eye tracking: A comprehensive guide to methods and measures}.
\newblock Oxford University Press, 2011.

\bibitem{hou2024unveiling}
B.~J. Hou, Y.~Abdrabou, F.~Weidner, and H.~Gellersen.
\newblock Unveiling variations: A comparative study of vr headsets regarding eye tracking volume, gaze accuracy, and precision.
\newblock In {\em 2024 IEEE Conference on Virtual Reality and 3D User Interfaces Abstracts and Workshops (VRW)}, pp. 650--655. IEEE, 2024.

\bibitem{hou2020multisensory}
H.~Hou and Y.~Gu.
\newblock Multisensory integration for self-motion perception.
\newblock {\em The senses: a comprehensive reference}, pp. 458--82, 2020.

\bibitem{huang2017reduced}
C.-K. Huang, J.-H. Chien, and K.-C. Siu.
\newblock The reduced lighting environment impacts gait characteristics during walking.
\newblock {\em International Journal of Industrial Ergonomics}, 61:126--130, 2017.

\bibitem{hutton2018individualized}
C.~Hutton, S.~Ziccardi, J.~Medina, and E.~S. Rosenberg.
\newblock Individualized calibration of rotation gain thresholds for redirected walking.
\newblock In {\em ICAT-EGVE}, pp. 61--64, 2018.

\bibitem{islam2019measuring}
A.~Islam, J.~Ma, T.~Gedeon, M.~Z. Hossain, and Y.-H. Liu.
\newblock Measuring user responses to driving simulators: A galvanic skin response based study.
\newblock In {\em 2019 IEEE International Conference on Artificial Intelligence and Virtual Reality (AIVR)}, pp. 33--337. IEEE, 2019.

\bibitem{ivy1929physiology}
A.~Ivy.
\newblock The physiology of vestibular nystagmus.
\newblock {\em Archives of Otolaryngology}, 9(2):123--134, 1929.

\bibitem{janeh2017walking}
O.~Janeh, E.~Langbehn, F.~Steinicke, G.~Bruder, A.~Gulberti, and M.~Poetter-Nerger.
\newblock Walking in virtual reality: Effects of manipulated visual self-motion on walking biomechanics.
\newblock {\em ACM Transactions on Applied Perception (TAP)}, 14(2):1--15, 2017.

\bibitem{kelly2008visual}
J.~W. Kelly, B.~Riecke, J.~M. Loomis, and A.~C. Beall.
\newblock Visual control of posture in real and virtual environments.
\newblock {\em Perception \& psychophysics}, 70:158--165, 2008.

\bibitem{kennedy1993simulator}
R.~S. Kennedy, N.~E. Lane, K.~S. Berbaum, and M.~G. Lilienthal.
\newblock Simulator sickness questionnaire: An enhanced method for quantifying simulator sickness.
\newblock {\em The international journal of aviation psychology}, 3(3):203--220, 1993.

\bibitem{kinsella2006effects}
J.~M. Kinsella-Shaw, S.~J. Harrison, C.~Colon-Semenza, and M.~T. Turvey.
\newblock Effects of visual environment on quiet standing by young and old adults.
\newblock {\em Journal of Motor Behavior}, 38(4):251--264, 2006.

\bibitem{krokos2022quantifying}
E.~Krokos and A.~Varshney.
\newblock Quantifying vr cybersickness using eeg.
\newblock {\em Virtual Reality}, 26(1):77--89, 2022.

\bibitem{langbehn2018redirected}
E.~Langbehn and F.~Steinicke.
\newblock Redirected walking in virtual reality.
\newblock {\em Encyclopedia of Computer Graphics and Games. Springer International Publishing}, 2018.

\bibitem{lappe1999perception}
M.~Lappe, F.~Bremmer, and A.~V. van~den Berg.
\newblock Perception of self-motion from visual flow.
\newblock {\em Trends in cognitive sciences}, 3(9):329--336, 1999.

\bibitem{lencer2019smooth}
R.~Lencer, A.~Sprenger, and P.~Trillenberg.
\newblock Smooth eye movements in humans: smooth pursuit, optokinetic nystagmus and vestibular ocular reflex.
\newblock {\em Eye movement research: An introduction to its scientific foundations and applications}, pp. 117--163, 2019.

\bibitem{lutwak2023user}
H.~Lutwak, T.~S. Murdison, and K.~W. Rio.
\newblock User self-motion modulates the perceptibility of jitter for world-locked objects in augmented reality.
\newblock In {\em 2023 IEEE International Symposium on Mixed and Augmented Reality (ISMAR)}, pp. 346--355. IEEE, 2023.

\bibitem{matsuda2022realistic}
N.~Matsuda, A.~Chapiro, Y.~Zhao, C.~Smith, R.~Bachy, and D.~Lanman.
\newblock Realistic luminance in vr.
\newblock In {\em SIGGRAPH Asia 2022 Conference Papers}, pp. 1--8, 2022.

\bibitem{matsuda2022hdr}
N.~Matsuda, Y.~Zhao, A.~Chapiro, C.~Smith, and D.~Lanman.
\newblock Hdr vr.
\newblock In {\em ACM SIGGRAPH 2022 Emerging Technologies}, pp. 1--2. 2022.

\bibitem{matsumoto2018biomechanical}
K.~Matsumoto, A.~Yamada, A.~Nakamura, Y.~Uchmura, K.~Kawai, and T.~Tanikawa.
\newblock Biomechanical parameters under curvature gains and bending gains in redirected walking.
\newblock In {\em 2018 IEEE Conference on Virtual Reality and 3D User Interfaces (VR)}, pp. 631--632. IEEE, 2018.

\bibitem{matthis2022retinal}
J.~S. Matthis, K.~S. Muller, K.~L. Bonnen, and M.~M. Hayhoe.
\newblock Retinal optic flow during natural locomotion.
\newblock {\em PLOS Computational Biology}, 18(2):e1009575, 2022.

\bibitem{meehan2002physiological}
M.~Meehan, B.~Insko, M.~Whitton, and F.~P. Brooks~Jr.
\newblock Physiological measures of presence in stressful virtual environments.
\newblock {\em Acm transactions on graphics (tog)}, 21(3):645--652, 2002.

\bibitem{mostajeran2024analyzing}
F.~Mostajeran, S.~Schneider, G.~Bruder, S.~K{\"u}hn, and F.~Steinicke.
\newblock Analyzing cognitive demands and detection thresholds for redirected walking in immersive forest and urban environments.
\newblock In {\em 2024 IEEE Conference Virtual Reality and 3D User Interfaces (VR)}, pp. 61--71. IEEE, 2024.

\bibitem{muller2023retinal}
K.~S. Muller, J.~Matthis, K.~Bonnen, L.~K. Cormack, A.~C. Huk, and M.~Hayhoe.
\newblock Retinal motion statistics during natural locomotion.
\newblock {\em Elife}, 12:e82410, 2023.

\bibitem{murata2004effects}
A.~Murata.
\newblock Effects of duration of immersion in a virtual reality environment on postural stability.
\newblock {\em International Journal of Human-Computer Interaction}, 17(4):463--477, 2004.

\bibitem{naaman2023young}
T.~Naaman, R.~Hayek, I.~Gutman, and S.~Springer.
\newblock Young, but not in the dark—the influence of reduced lighting on gait stability in middle-aged adults.
\newblock {\em Plos one}, 18(5):e0280535, 2023.

\bibitem{nakamura2013effects}
S.~Nakamura, T.~Seno, H.~Ito, and S.~Sunaga.
\newblock Effects of dynamic luminance modulation on visually induced self-motion perception: observers' perception of illumination is important in perceiving self-motion.
\newblock {\em Perception}, 42(2):153--162, 2013.

\bibitem{neth2012velocity}
C.~T. Neth, J.~L. Souman, D.~Engel, U.~Kloos, H.~H. Bulthoff, and B.~J. Mohler.
\newblock Velocity-dependent dynamic curvature gain for redirected walking.
\newblock {\em IEEE transactions on visualization and computer graphics}, 18(7):1041--1052, 2012.

\bibitem{nguyen2020effect}
A.~Nguyen, Y.~Rothacher, E.~Efthymiou, B.~Lenggenhager, P.~Brugger, L.~Imbach, and A.~Kunz.
\newblock Effect of cognitive load on curvature redirected walking thresholds.
\newblock In {\em 26th ACM Symposium on Virtual Reality Software and Technology}, pp. 1--5, 2020.

\bibitem{nguyen2018individual}
A.~Nguyen, Y.~Rothacher, B.~Lenggenhager, P.~Brugger, and A.~Kunz.
\newblock Individual differences and impact of gender on curvature redirection thresholds.
\newblock In {\em Proceedings of the 15th acm symposium on applied perception}, pp. 1--4, 2018.

\bibitem{nguyen2020effect_emobdiment}
A.~Nguyen, Y.~Rothacher, B.~Lenggenhager, P.~Brugger, and A.~Kunz.
\newblock Effect of sense of embodiment on curvature redirected walking thresholds.
\newblock In {\em ACM Symposium on Applied Perception 2020}, pp. 1--5, 2020.

\bibitem{nilsson201815}
N.~C. Nilsson, T.~Peck, G.~Bruder, E.~Hodgson, S.~Serafin, M.~Whitton, F.~Steinicke, and E.~S. Rosenberg.
\newblock 15 years of research on redirected walking in immersive virtual environments.
\newblock {\em IEEE computer graphics and applications}, 38(2):44--56, 2018.

\bibitem{orlosky2019using}
J.~Orlosky, B.~Huynh, and T.~Hollerer.
\newblock Using eye tracked virtual reality to classify understanding of vocabulary in recall tasks.
\newblock In {\em 2019 IEEE International Conference on Artificial Intelligence and Virtual Reality (AIVR)}, pp. 66--667. IEEE, 2019.

\bibitem{owen2021adaptive}
L.~Owen, J.~Browder, B.~Letham, G.~Stocek, C.~Tymms, and M.~Shvartsman.
\newblock Adaptive nonparametric psychophysics.
\newblock {\em arXiv preprint arXiv:2104.09549}, 2021.

\bibitem{peck2021divrsify}
T.~C. Peck, K.~A. McMullen, and J.~Quarles.
\newblock Divrsify: Break the cycle and develop vr for everyone.
\newblock {\em IEEE Computer Graphics and Applications}, 41(6):133--142, 2021.

\bibitem{peck2020mind}
T.~C. Peck, L.~E. Sockol, and S.~M. Hancock.
\newblock Mind the gap: The underrepresentation of female participants and authors in virtual reality research.
\newblock {\em IEEE transactions on visualization and computer graphics}, 26(5):1945--1954, 2020.

\bibitem{prokop1997visual}
T.~Prokop, M.~Schubert, and W.~Berger.
\newblock Visual influence on human locomotion modulation to changes in optic flow: Modulation to changes in optic flow.
\newblock {\em Experimental brain research}, 114:63--70, 1997.

\bibitem{ranti2020blink}
C.~Ranti, W.~Jones, A.~Klin, and S.~Shultz.
\newblock Blink rate patterns provide a reliable measure of individual engagement with scene content.
\newblock {\em Scientific reports}, 10(1):8267, 2020.

\bibitem{razzaque2005redirected}
S.~Razzaque.
\newblock {\em Redirected walking}.
\newblock The University of North Carolina at Chapel Hill, 2005.

\bibitem{razzaque2001redirected}
S.~Razzaque, Z.~Kohn, and M.~C. Whitton.
\newblock Redirected walking.
\newblock In {\em Proceedings of EUROGRAPHICS}, vol.~9, pp. 105--106. Citeseer, 2001.

\bibitem{riccio1991ecological}
G.~E. Riccio and T.~A. Stoffregen.
\newblock An ecological theory of motion sickness and postural instability.
\newblock {\em Ecological psychology}, 3(3):195--240, 1991.

\bibitem{rietzler2018rethinking}
M.~Rietzler, J.~Gugenheimer, T.~Hirzle, M.~Deubzer, E.~Langbehn, and E.~Rukzio.
\newblock Rethinking redirected walking: On the use of curvature gains beyond perceptual limitations and revisiting bending gains.
\newblock In {\em 2018 IEEE International Symposium on Mixed and Augmented Reality (ISMAR)}, pp. 115--122. IEEE, 2018.

\bibitem{rugelj2014influence}
D.~Rugelj, G.~Gomi{\v{s}}{\v{c}}ek, and F.~Sev{\v{s}}ek.
\newblock The influence of very low illumination on the postural sway of young and elderly adults.
\newblock {\em PLoS One}, 9(8):e103903, 2014.

\bibitem{saeedpour2024perceptual}
M.~R. Saeedpour-Parizi, N.~L. Williams, T.~Wong, P.~Guan, D.~Manocha, and I.~M. Erkelens.
\newblock Perceptual {T}hresholds for {R}adial {O}ptic {F}low {D}istortion in {N}ear-{E}ye {S}tereoscopic {D}isplays.
\newblock {\em IEEE {T}ransactions on {V}isualization and {C}omputer {G}raphics}, 30(5):2570--2579, 2024.

\bibitem{sara2017effect}
G.~Sara, P.~Tony, B.~Roberto, H.~Kerstin, and B.~Mariagrazia.
\newblock The effect of luminance condition on form, form-from-motion and motion perception.
\newblock {\em pathways}, 8:11, 2017.

\bibitem{huang2023simple}
D.~S. Schwarzkopf and Z.~Huang.
\newblock A simple statistical framework for small sample studies.
\newblock {\em Psychological methods}, December 2024.

\bibitem{schweigart1997gaze}
G.~Schweigart, T.~Mergner, I.~Evdokimidis, S.~Morand, and W.~Becker.
\newblock Gaze stabilization by optokinetic reflex (okr) and vestibulo-ocular reflex (vor) during active head rotation in man.
\newblock {\em Vision research}, 37(12):1643--1652, 1997.

\bibitem{shayman2022multisensory}
C.~S. Shayman, J.~K. Stefanucci, P.~C. Fino, and S.~H. Creem-Regehr.
\newblock Multisensory cue combination during navigation: lessons learned from replication in real and virtual environments.
\newblock In {\em 2022 IEEE International Symposium on Mixed and Augmented Reality Adjunct (ISMAR-Adjunct)}, pp. 276--277. IEEE, 2022.

\bibitem{siegler2000self}
I.~Siegler, I.~Viaud-Delmon, I.~Israel, and A.~Berthoz.
\newblock Self-motion perception during a sequence of whole-body rotations in darkness.
\newblock {\em Experimental brain research}, 134(1):66--73, 2000.

\bibitem{smith2018small}
P.~L. Smith and D.~R. Little.
\newblock Small is beautiful: In defense of the small-n design.
\newblock {\em Psychonomic bulletin \& review}, 25:2083--2101, 2018.

\bibitem{soffel2016postural}
F.~Soffel, M.~Zank, and A.~Kunz.
\newblock Postural stability analysis in virtual reality using the htc vive.
\newblock In {\em Proceedings of the 22nd ACM Conference on Virtual Reality Software and Technology}, pp. 351--352, 2016.

\bibitem{steinicke2009estimation}
F.~Steinicke, G.~Bruder, J.~Jerald, H.~Frenz, and M.~Lappe.
\newblock Estimation of detection thresholds for redirected walking techniques.
\newblock {\em IEEE transactions on visualization and computer graphics}, 16(1):17--27, 2009.

\bibitem{steinicke2013human}
F.~Steinicke, Y.~Visell, J.~Campos, and A.~L{\'e}cuyer.
\newblock {\em Human walking in virtual environments}, vol.~56.
\newblock Springer, 2013.

\bibitem{stoffregen2008motion}
T.~A. Stoffregen, E.~Faugloire, K.~Yoshida, M.~B. Flanagan, and O.~Merhi.
\newblock Motion sickness and postural sway in console video games.
\newblock {\em Human factors}, 50(2):322--331, 2008.

\bibitem{stoffregen1998postural}
T.~A. Stoffregen and L.~J. Smart~Jr.
\newblock Postural instability precedes motion sickness.
\newblock {\em Brain research bulletin}, 47(5):437--448, 1998.

\bibitem{stoffregen2010stance}
T.~A. Stoffregen, K.~Yoshida, S.~Villard, L.~Scibora, and B.~G. Bardy.
\newblock Stance width influences postural stability and motion sickness.
\newblock {\em Ecological Psychology}, 22(3):169--191, 2010.

\bibitem{sutherland2023practical}
C.~Sutherland, D.~Hare, P.~J. Johnson, D.~W. Linden, R.~A. Montgomery, and E.~Droge.
\newblock Practical advice on variable selection and reporting using akaike information criterion.
\newblock {\em Proceedings of the Royal Society B}, 290(2007):20231261, 2023.

\bibitem{takeuchi2000velocity}
T.~Takeuchi and K.~K. De~Valois.
\newblock Velocity discrimination in scotopic vision.
\newblock {\em Vision research}, 40(15):2011--2024, 2000.

\bibitem{tanahashi2007effects}
S.~Tanahashi, H.~Ujike, R.~Kozawa, and K.~Ukai.
\newblock Effects of visually simulated roll motion on vection and postural stabilization.
\newblock {\em Journal of neuroengineering and rehabilitation}, 4:1--11, 2007.

\bibitem{tanguy2008vestibulo}
S.~Tanguy, G.~Quarck, O.~Etard, A.~Gauthier, and P.~Denise.
\newblock Vestibulo-ocular reflex and motion sickness in figure skaters.
\newblock {\em European journal of applied physiology}, 104:1031--1037, 2008.

\bibitem{taylor1967pest}
M.~Taylor and C.~D. Creelman.
\newblock Pest: Efficient estimates on probability functions.
\newblock {\em The Journal of the Acoustical Society of America}, 41(4A):782--787, 1967.

\bibitem{tirado2019analysis}
C.~A. Tirado~Cortes, H.-T. Chen, and C.-T. Lin.
\newblock Analysis of vr sickness and gait parameters during non-isometric virtual walking with large translational gain.
\newblock In {\em Proceedings of the 17th International Conference on Virtual-Reality Continuum and its Applications in Industry}, pp. 1--10, 2019.

\bibitem{ujike2004effects}
H.~Ujike, T.~Yokoi, and S.~Saida.
\newblock Effects of virtual body motion on visually-induced motion sickness.
\newblock In {\em The 26th Annual International Conference of the IEEE Engineering in Medicine and Biology Society}, vol.~1, pp. 2399--2402. IEEE, 2004.

\bibitem{van2010eye}
D.~C. Van~Barneveld and A.~John Van~Opstal.
\newblock Eye position determines audiovestibular integration during whole-body rotation.
\newblock {\em European Journal of Neuroscience}, 31(5):920--930, 2010.

\bibitem{van2011defining}
R.~Van~der Lans, M.~Wedel, and R.~Pieters.
\newblock Defining eye-fixation sequences across individuals and tasks: the binocular-individual threshold (bit) algorithm.
\newblock {\em Behavior research methods}, 43:239--257, 2011.

\bibitem{vrieze2012model}
S.~I. Vrieze.
\newblock Model selection and psychological theory: a discussion of the differences between the akaike information criterion (aic) and the bayesian information criterion (bic).
\newblock {\em Psychological methods}, 17(2):228, 2012.

\bibitem{warren2001optic}
W.~H. Warren~Jr, B.~A. Kay, W.~D. Zosh, A.~P. Duchon, and S.~Sahuc.
\newblock Optic flow is used to control human walking.
\newblock {\em Nature neuroscience}, 4(2):213, 2001.

\bibitem{warwick1998evaluating}
L.~Warwick-Evans, N.~Symons, T.~Fitch, and L.~Burrows.
\newblock Evaluating sensory conflict and postural instability. theories of motion sickness.
\newblock {\em Brain research bulletin}, 47(5):465--469, 1998.

\bibitem{watson1983quest}
A.~B. Watson and D.~G. Pelli.
\newblock Quest: A bayesian adaptive psychometric method.
\newblock {\em Perception \& psychophysics}, 33(2):113--120, 1983.

\bibitem{wichmann2001psychometric1}
F.~A. Wichmann and N.~J. Hill.
\newblock The psychometric function: I. {F}itting, sampling, and goodness of fit.
\newblock {\em Perception \& psychophysics}, 63(8):1293--1313, 2001.

\bibitem{wichmann2001psychometric2}
F.~A. Wichmann and N.~J. Hill.
\newblock The psychometric function: I{I}. {B}ootstrap-based confidence intervals and sampling.
\newblock {\em Perception \& psychophysics}, 63:1314--1329, 2001.

\bibitem{williams2019estimation}
N.~L. Williams and T.~C. Peck.
\newblock Estimation of rotation gain thresholds considering fov, gender, and distractors.
\newblock {\em IEEE transactions on visualization and computer graphics}, 25(11):3158--3168, 2019.

\bibitem{yang2024eye}
B.~Yang, J.~Intoy, and M.~Rucci.
\newblock Eye blinks as a visual processing stage.
\newblock {\em Proceedings of the National Academy of Sciences}, 121(15):e2310291121, 2024.

\bibitem{yoshimoto2016motion}
S.~Yoshimoto, K.~Okajima, and T.~Takeuchi.
\newblock Motion perception under mesopic vision.
\newblock {\em Journal of Vision}, 16(1):16--16, 2016.

\bibitem{zhao2023retinal}
Y.~Zhao, D.~Lindberg, B.~Cleary, O.~Mercier, R.~Mcclelland, E.~Penner, Y.-J. Lin, J.~Majors, and D.~Lanman.
\newblock Retinal-resolution varifocal vr.
\newblock In {\em ACM SIGGRAPH 2023 Emerging Technologies}, pp. 1--3. 2023.

\end{thebibliography}
